\documentclass[aps,a4paper,pre,showpacs,floats,nofootinbib,superscriptaddress,twocolumn]{revtex4}
\usepackage{latexsym,amsmath}
\usepackage{amsbsy}
\usepackage[dvips,usenames]{color}
\usepackage{graphicx}
\usepackage{amssymb}
\usepackage{color}

\newcommand{\avk}{\langle k \rangle}
\newcommand{\fluck}[1]{\langle {#1}^2\rangle}
\newcommand{\nper}{{\nu_\perp}}

\newcommand{\npar}{{\nu_\parallel}}

\newcommand{\kmax}{{k_{\mathrm{max}}}}

\begin{document}

\title{Langevin approach for the dynamics of the contact process on
  annealed scale-free networks}

\author{Mari{\'a}n Bogu{\~n}{\'a}}

\affiliation{Departament de F{\'\i}sica Fonamental, Universitat de
  Barcelona, Mart\'{\i} i Franqu\`es 1, 08028 Barcelona, Spain}

\author{Claudio Castellano}
\affiliation{SMC, INFM-CNR and
Dipartimento di Fisica, ``Sapienza'' Universit\`a di Roma,
P.le Aldo Moro 2, I-00185 Roma, Italy}

\author{Romualdo Pastor-Satorras}
\affiliation{Departament de F\'\i sica i Enginyeria Nuclear, Universitat
  Polit\`ecnica de Catalunya, Campus Nord B4, 08034 Barcelona, Spain}

\date{\today}

\begin{abstract}
  We study the dynamics of the
  contact-process, one of the simplest nonequilibrium stochastic
  processes, taking place on a scale-free network.
    We consider the network topology as annealed, i.e. all links are
  rewired at each microscopic time step, so that no dynamical
  correlation can build up. This is a practical implementation of the
  absence of correlations assumed by mean-field approaches. We present
  a detailed analysis of the contact process in terms of a Langevin
  equation, including explicitly the effects of stochastic
  fluctuations in the number of particles in finite networks.
  This allows us to determine analytically the survival time for
  spreading experiments and the density of active sites in surviving
  runs.  The fluctuations in the topological structure induce
  anomalous scaling effects with respect to the system size when the
  degree distribution has an ``hard'' upper bound.  When the upper
  bound is soft, the presence of outliers with huge connectivity
  perturbs the picture even more, inducing an apparent shift of the
  critical point.  In light of these findings, recent theoretical and
  numerical results in the literature are critically reviewed.

\end{abstract}


\maketitle

\section{Introduction}

The study of the effects of an heterogeneous topology on equilibrium
and nonequilibrium dynamical processes has lately experienced an
active interest from the statistical physics community
\cite{dorogovtsev07:_critic_phenom}. Indeed, it has been observed in
recent years that many natural and man-made systems are well
characterized in terms of complex networks or graphs
\cite{Albert:2002,Dorogovtsev:2003}, in which vertices represent
elementary units in the system, while edges stand for pairwise
interactions between elements. Most real networked systems can be
characterized by a heterogenous complex topology, showing remarkable
universal features, such as the small world property \cite{Watts:1998}
and a scale-free connectivity pattern \cite{Barabasi:1999}.  The
small-world property refers to the fact that the average distance
$\langle \ell \rangle $ between any two vertices---defined as the
smallest number of edges on a path between one and the other---is very
small, scaling logarithmically or even more slowly with the network
size $N$ \cite{havlin03}. This is to be compared to the power-law
scaling $\langle \ell \rangle \sim N^{1/d} $ in a $d$-dimensional
lattice. Since the logarithm grows slower than any power-law function,
even if $d$ is very large, small-world networks can be thought of as
highly compact objects of infinite dimensionality. On the other hand,
scale-free (SF) networks are typically characterized by a degree
distribution $P(k)$, defined as the probability that a randomly
selected vertex has degree $k$---is connected to $k$ other
vertices---that decreases as a power-law,
\begin{equation}
  P(k) \sim k^{-\gamma},
\end{equation}
where $\gamma$ is a characteristic degree exponent, usually in the
range $2 < \gamma \leq 3$ \cite{Albert:2002,Dorogovtsev:2003}.

Dynamical processes taking place on top of complex networks arise in a
wide variety of scientific and technological contexts. For example, we
can mention the transmission of information packets on the Internet
\cite{RomusVespasbook}, the spreading of biological diseases on social
networks or computer viruses in computer infrastructures
\cite{Vespignani:2001,Lloyd:2001}, etc.  The interest in the study of
these dynamics was triggered by the observation that the heterogeneous
connectivity pattern observed in SF networks with diverging degree
fluctuations can lead to very surprising outcomes, such as an extreme
weakness in the face of targeted attacks aimed at destroying the most
connected vertices \cite{havlin01,newman00}, or the ease of
propagation of infective agents \cite{Romualdo:2001,Lloyd:2001}. These
properties are due to the critical interplay between topology and
dynamics in heterogeneous networks and are absent in their homogeneous
counterparts.  After those initial discoveries, a real avalanche of
new results have been put forward, including classical equilibrium
systems \cite{isingvespi,isingmendes,dorogovtsev04:_potts} and
non-equilibrium processes such as epidemic spreading
\cite{Romualdo:2001,Lloyd:2001}, reaction-diffusion processes
\cite{originalA+A,Catanzaro:2005b,v.07:_react} and dynamics with
absorbing states \cite{Castellano:2006,Hong:2007}.  For an extensive
review of recent results we refer the reader to
Ref.~\cite{dorogovtsev07:_critic_phenom}.

The analytical approach to the study of dynamical processes on complex
networks is dominated by the application of heterogeneous mean-field
theory \cite{dorogovtsev07:_critic_phenom}. Heterogeneous mean-field
theory (HMF) is based in two basic assumptions: (i) the homogeneous
mixing hypothesis, stating that all vertices with the same degree
(within the same degree class) share the same dynamical properties;
and (ii) the assumption that fluctuations are not relevant, and
therefore analytical studies can be conducted within a deterministic
approach. This last fact is in some sense natural, since the
small-world property implies that dynamical fluctuations in a network
are so close together that they can be washed away in very few time
steps\footnote{At variance with what happens in regular lattices below
  the critical dimension, where in particular, close to a critical
  point, dynamics is governed by fluctuations~\cite{goldenfeld}.}. HMF
has proved to be extremely useful in providing a very accurate
description of the behavior of most dynamical processes on complex
networks \cite{dorogovtsev07:_critic_phenom}. On the other hand, in
other instances, such as in the nonequilibrium contact process (CP)
\cite{Marro} a debate has arisen about the comparison between
numerical simulations on SF networks and HMF predictions
\cite{Castellano:2006,hong07:_comment,castellano07:_reply}.
Underlying this controversy is the fact that while the HMF approach
considers all relevant quantities as deterministic, and hence assumes
an infinite system size, numerical simulations are performed on finite
systems and thus are necessarily influenced by stochastic fluctuations
due to the finite number of particles, in particular close to an
absorbing state phase transition \cite{Marro}.  Finite size effects
are very strong in networks and an appropriate theoretical framework
for them is necessary to compare simulations with HMF results.  For
this reason, in Ref.~\cite{Castellano:2008} the CP was considered on
the simplest network substrate, which is an annealed network, in which
the quenched disorder imposed by the actual connections in the network
is not considered. In this scenario, it was possible to deduce, by
means of qualitative arguments, the correct size scaling of the CP in
this kind of networks, in very good agreement with numerical
simulations.

In this paper we present a more detailed analysis of the CP in SF
networks, deriving the corresponding Langevin equation describing its
dynamics for the case of annealed networks. The analysis of this
equation allows us to uncover the correct finite size scaling behavior
of the CP in heterogeneous networks, providing the exact value of the
critical exponents describing this system. Surprisingly, the critical
behavior of the CP turns out to be extremely sensitive to the
particular degree cutoff chosen for the construction of the network,
in agreement with previous results obtained from a more
phenomenological approach \cite{Castellano:2008}. In particular,
critical exponents depend explicitly on the way the degree cutoff
diverges with the system size, if it scales sufficiently slowly.  If the
scaling is instead fast, additional complications arise and
fluctuations of the degree distribution strongly perturb the picture.

We have organized our paper as follows. In
Sec.~\ref{sec:annealed-scale-free} we describe the main properties of
annealed networks, which represent the simplest network substrate for
a dynamical process, in which mean-field theory is supposed to be
exact. We focus, in particular, on the effects of the maximum degree
allowed on the network cutoff and on its fluctuations.
Sec.~\ref{sec:cont-proc-compl} defines the CP on complex
networks, whose mean-field analysis is reviewed in
Sec.~\ref{sec:heter-mean-field}. In
Sec.~\ref{sec:finite-size-scaling} we comment on the different
approaches followed in the past to deal with finite size effects on
the CP in SF networks. The general Langevin theory for the CP in
networks is presented in Sec.~\ref{sec:langevin-approach-cp}, while
Sec.~\ref{sec:CPSF} focuses on the analysis of annealed
networks. Sec.~\ref{sec:meaning-finite-size} discusses the meaning of
finite size effects and finite size scaling in heterogeneous
networks. In Sec.~\ref{sec:effect-outliers} we present a digression
to the case of annealed networks with outliers, that is, vertices with
a degree much larger than the average maximum degree expected in the
network.  Finally, we draw our conclusions in
Sec.~\ref{sec:conclusions}. Some technical questions are developed in
several Appendices.

\section{Annealed scale-free networks}
\label{sec:annealed-scale-free}

The topological properties of any complex network are fully encoded in
its adjacency matrix $a_{ij}$, taking the value $a_{ij}=1$ if there is
an edge connecting vertices $i$ and $j$, and zero otherwise. In the
so-called \textit{quenched networks}, the values of the adjacency
matrix are fixed in time.  For large quenched networks, a statistical
characterization in terms of the degree distribution $P(k)$ and the
degree correlations $P(k'|k)$, defined as the conditional probability that
a vertex of degree $k$ is connected to a vertex of degree
$k'$~\cite{Pastor-Satorras:2001,serrano07:_correl}, is useful as a
compact way to express the essential features of the adjacency
matrix\footnote{A more detailed characterization can be made using
  higher order degree correlations, see
  Ref.~\cite{serrano07:_correl}.}.
Quenched networks are the typical output of most network models,
such as the configuration model (CM)
\cite{bekessi72,benderoriginal,bollobas1980,molloy95}, the
uncorrelated configuration model \cite{Catanzaro:2005}, the class of
models with hidden variables \cite{Boguna:2003b}, linear preferential
attachment models \cite{Barabasi:1999,mendes99}, etc.  In this case,
each network must be considered as a representative of a statistical
ensemble of random networks, which is characterized by the $P(k)$ and
$P(k'|k)$ probability distributions.  When a dynamical process takes
place on top of such a network, one is considering the network as
frozen, with respect to the characteristic time scale $\tau_D$ of the
dynamics.  In this case, in a numerical analysis of a dynamical
process, one must consider the dynamics over many different quenched
networks, all belonging to the network ensemble with the same
statistically equivalent topological properties, and perform an
ensemble average to compute the average dynamical quantities.

In other instances, on the other hand, the very network is a dynamical
object, changing in time over a certain time scale $\tau_N$. In this
case, the correct topological characterization is strictly
statistical, given in terms of the degree distribution $P(k)$ and
the degree correlations $P(k'|k)$.  In the limit $\tau_N \ll
\tau_D$, that is, when the network connections are completely
reshuffled between any two microscopic steps of the dynamics, while
keeping fixed $P(k)$ and $P(k'|k)$, the resulting networks are called
\textit{annealed}
\cite{gil05:_optim_disor,stauffer_annealed2005,weber07:_gener}.  Apart
from the cases where they describe the actual evolution of real
systems, annealed networks are extremely important from a theoretical
point of view, because mean-field predictions for dynamical processes
on networks are usually obtained in this limit, via the so-called
annealed network approximation~\cite{dorogovtsev07:_critic_phenom}.
In practice one replaces the adjacency matrix $a_{ij}$ by its ensemble
average $\bar{a}(k_i,k_j)$, defining the probability that two vertices
of degree $k_i$ and $k_j$ are connected.  This average is given by
\begin{equation}
  \bar{a}(k,k') = \frac{1}{N P(k)}\frac{1}{N P(k')} \sum_{i\in
    k}\sum_{j\in k'} 
  a_{ij} \equiv \frac{k' P(k|k')}{N P(k)},
  \label{eq:3}
\end{equation}
where notation $i \in k$ means summation for all vertices of degree
$k$.  Taking the case of uncorrelated networks, with $P(k|k')=k
P(k)/\avk$ \cite{Dorogovtsev:2002}, the simple form $ \bar{a}(k,k')= k
k' / N \avk$ results.

From a numerical point of view, the simulation of dynamics on annealed
networks implies the re-generation of the whole network every time a
microscopic dynamic step is performed~\cite{weber07:_gener}. For
uncorrelated networks this can be efficiently implemented in CP-like
dynamics. In this case, an annealed network of size $N$ is completely
defined by its degree sequence $\{k_1, \ldots, k_N\}$, where the
degrees $k_i$ are integer random numbers, extracted according to the
degree distribution $P(k)$, and restricted between a lower bound $m$
and an upper bound $M \leq N$.  Degree correlations are given by
$P(k'|k)=k' P(k')/\avk$. Thus, every time we need to find a nearest
neighbor of a vertex, it is selected at random with probability $k'
P(k')/\avk$ among the $N$ vertices present in the network.

Finite SF networks are additionally characterized by another
parameter, the degree cutoff $k_c(N)$~\cite{Dorogovtsev:2002}, that
is the average value of the actual maximum degree $k_{\mathrm{max}}$
in a single realization of the degree sequence: $k_c(N)=\langle
k_{\mathrm{max}} \rangle$.  In general, $k_c$ is a non decreasing
function of the network size and, as we shall see below, the CP
dynamics is very sensitive to its actual size dependence.

Notice that the value of the cutoff plays a relevant role in the
determination of degree correlations in finite quenched
networks~\cite{Boguna:2004}.
It is known that for the network to be closed without degree-degree
correlations and no multiple edges or self-loops one must
impose that degrees are smaller than the {\em structural}
cutoff $\sim N^{1/2}$~\cite{Catanzaro:2005}.
In uncorrelated annealed networks, however, since they are
by construction uncorrelated, such a restriction does not apply, and
any cutoff is in principle possible.

Simple considerations based on extreme value theory \cite{Boguna:2004}
give the probability distribution $P_\mathrm{max}(k_\mathrm{max})$ of
observing a maximum degree $k_{\mathrm{max}}$ among $N$ degrees
independently sampled from a distribution $P(k) \sim k^{-\gamma}$ and
bounded by the constraint $m \le k \le M$. In the continuous degree
approximation, the distribution of maximum degrees takes the form
\begin{equation}
  P_{\mathrm{max}}(k_{\mathrm{max}}=k) =
  N(\gamma-1)\frac{(m^{1-\gamma}-k^{1-\gamma})^{N-1}}
  {(m^{1-\gamma}-M^{1-\gamma})^N} k^{-\gamma}   
\end{equation}
Using this expression one can compute explicitly the value of
$k_c(N)$,
obtaining two different behaviors, depending on whether
$M/m$ is larger or smaller than $N^{1/(\gamma-1)}$, namely
\begin{equation}
  k_c(N) =
  \left\{ 
    \begin{array}{lr}
      \displaystyle{M}, & \frac{M}{m} \ll
      N^{1/(\gamma-1)} \\  
      \displaystyle{m \Gamma\left(\frac{\gamma-2}{\gamma-1} \right)
                    N^{1/(\gamma-1)} },  & \frac{M}{m} \gg
                  N^{1/(\gamma-1)} 
    \end{array} 
  \right. ,
  \label{eq:5}
\end{equation}
where $\Gamma(z)$ is the Gamma function \cite{abramovitz}. 

For the network to be SF the upper bound of the degree
distribution must diverge with the system size: $M \sim N^{1/\omega}$.
The parameter $\omega \ge 0$ is in principle arbitrary, but its value
strongly affects the nature of the actual maximum of the degree
sequence.  If $M$ diverges not faster than $N^{1/(\gamma-1)}$
(i. e. $\omega \ge \gamma-1$), then $k_c=M$ is a hard cutoff, with no
degree larger than $k_c$.  For $ \omega < \gamma-1$ instead, $k_c \sim
N^{1/(\gamma-1)}$ is just an average cutoff, but
$\langle \kmax^2 \rangle$ grows with the network size as $N^{1+(3-\gamma)/\omega} \gg 
k_c^2$, indicating that fluctuations diverge.

As a consequence, the maximum degree present in the
degree sequence has wide fluctuations and {\em outliers}, i.e. nodes
with a degree much larger than $k_c$, may be present in the network.
It is important to stress that taking $\omega=\gamma-1$ is very
different from setting $M=\infty$ from the beginning ($\omega=0$) or
$M=N$ ($\omega=1$), as it is usually done in the quenched
configuration model \cite{molloy95}.  In both cases the average cutoff
$k_c$ scales as $N^{1/(\gamma-1)}$, but for $\omega=\gamma-1$ this is
a hard cutoff and fluctuations of the value of $\kmax$ around $k_c$
are bounded.

The presence of outliers and large fluctuations in the maximum degree
has a strong effect on the dynamics on annealed networks. In particular,
as we will see, the relevant quantity characterizing the size effects
on the dynamics is the second moment of the degree distribution
\begin{equation}
  g= \frac{ \fluck{k}}{\avk^2}.
  \label{eq:10}
\end{equation}
The fluctuating nature of this quantity from sample to sample
can be assessed by looking at
its standard deviation $\sigma_g$, that can be easily computed, given
the uncorrelated nature of the degrees in annealed networks. Thus, we
have the relative fluctuations
\begin{equation}
  \frac{\sigma_g^2}{g^2} = \frac{1}{N}
  \left(\frac{\langle k^4\rangle}{\langle k^2\rangle^2} -1 \right).
\end{equation}
Assuming that
$\langle k^n\rangle \sim \langle \kmax^{n+1-\gamma} \rangle$, we have
\begin{equation}
  \frac{\sigma_g^2}{g^2}
  \sim \frac{\langle \kmax^{5-\gamma} \rangle}
{N \langle \kmax^{3-\gamma} \rangle^2} \sim  \left\{ 
    \begin{array}{lr}
      \displaystyle{N^{2(3-\gamma)\left(\frac{1}{\omega} -\frac{1}{(\gamma-1)}\right)}}, & \mbox{for} \; \;
      \omega < \gamma-1 \\  
      \displaystyle{N^{(\gamma-1)/\omega -1}},  & \mbox{for} \; \;
      \omega \geq \gamma-1 
    \end{array} 
  \right. ,
\end{equation}
Thus, fluctuations vanish in the large size limit for $\omega \geq
\gamma-1$,  while for $\omega<\gamma-1$, the fluctuations of $g$
diverge as a power law with the network size $N$. 


In the rest of the paper, we will mainly discuss the simplest case
$\omega \ge \gamma-1$, considering often the cases $\omega=2$ and
$\omega=\gamma-1$. 
The more delicate issue of the effect of
outliers on the behavior of CP on SF networks will be touched
only in Sec.~\ref{sec:effect-outliers}.  Notice that the value $\omega=2$
has no special meaning here and it is just an example of what occurs
for $\omega>\gamma-1$. At odds with the case of quenched networks, the
structural cutoff $k_c=N^{1/2}$ does not play any role in annealed
networks.

\section{The contact process on complex networks}
\label{sec:cont-proc-compl}

We consider the contact process (CP)~\cite{Marro} on heterogeneous
networks, which is defined as follows \cite{Castellano:2006}. An
initial fraction $\rho_0$ of vertices is randomly chosen and occupied
by a particle.  Dynamics evolves in continuous time by the following
stochastic processes: Particles in vertices of degree $k$ create
offsprings into their nearest neighbors at rate $\lambda/k$,
independently of the degree $k'$ of the nearest neighbors. At the same
time, particles disappear at rate $\mu$ that, without loss of
generality, is set to $\mu=1$. From a computational point of view, the
CP can be efficiently implemented by means of a sequential updating
algorithm \cite{Marro,Castellano:2006}: At each time step $t$, a
particle in a vertex $i$ is chosen at random. With probability
$p=1/(\lambda+1)$ the particle disappears. On the other hand, with
probability $1-p=\lambda/(\lambda+1)$, the particle may generate an
offspring. In this case, a vertex $j$, nearest neighbor of $i$, is
selected at random. If $j$ is empty, a new particle is created of it;
otherwise, nothing happens. In any case, time is updated as $t \to
t+[(1+\lambda)n(t)]^{-1}$, where $n(t)$ is the number of particles at
the beginning of the time step. Notice that the factor $(1+\lambda)$
in the time update is due to the fact that each infected particle can
perform two independent actions, either infect a neighbor (at rate
$\lambda$) or become healthy again (at rate $\mu=1$). This factor was
neglected in previous implementations of the CP in complex networks
\cite{Castellano:2006,castellano07:_reply,Castellano:2008}. The
results of these works remain, however, unaltered, since the factor is
irrelevant for steady state properties and amounts only
to a rescaling for time dependent properties.

In Euclidean $d$-dimensional lattices, the CP undergoes a
nonequilibrium phase transition \cite{Marro} between an absorbing
state, with zero particle density, and an active phase, with average
constant density of particles, which takes place at a critical point
$\lambda_c$.  This phase transition is characterized in terms of the
order parameter $\rho$, defined as the average density of particles in
the steady state.  Defining $\Delta = \lambda -\lambda_c$, we observe
for $\Delta<0$, and in infinite lattices, an absorbing phase with
$\rho=0$. For $\Delta>0$, on the other hand, the system sets in an
active phase with a nonzero order parameter, obeying $\rho \sim
\Delta^\beta$. Close to the critical point, the system is also
characterized by diverging correlation length and time scales, namely
$\xi \sim |\Delta|^{-\nper}$ and $\tau \sim |\Delta|^{-\npar}$. The
critical exponents $\beta$, $\nper$ and $\npar$ characterize the
steady state properties of the transition. It is also possible to look
at the time dependent behavior at the critical point.  Thus, for
example, the particle density is observed to decay in time as $\rho(t)
\sim t^{-\theta}$. Different quantities can also be defined to
evaluate the time properties of spreading experiments, in which the
dynamics evolves starting from a single particle. In this case we can
define the survival probability, $S(t)$, as the probability that the
activity lasts longer that $t$, finding at the critical point $S(t)
\sim t^{-\delta}$. These and other critical exponents are not
independent, but are related by a set of scaling and hyperscaling
relations \cite{Marro}. Thus it is possible to give a full
characterization of the phase transition of CP in Euclidean lattices
using only three exponents, that we can take to be (without lack of
generality) $\beta$, $\nper$ and $\npar$.  Below the critical
dimension $d_c=4$, the exponents are nontrivial, and depend explicitly
on $d$.  For $d>d_c$, the exponents take the classical MF values
$\beta=\npar=1$, $\nper=1/2$.

\section{Heterogeneous mean-field theory for the CP}
\label{sec:heter-mean-field}

Heterogeneous mean-field theory (HMF) is the basic starting point to
obtain an analytical understanding of the behavior of any dynamical
process on a complex network~\cite{dorogovtsev07:_critic_phenom}.  In
order to take into account the possible fluctuations induced by the
network connectivity, the partial densities $\rho_k(t)$ of occupied
vertices of degree $k$~\cite{Vespignani:2001,marianproc} are
considered, from which the total density of particles is obtained as
$\rho(t) = \sum_k \rho_k (t) P(k)$.  In the spirit of standard
mean-field theories~\cite{stanley}, the fact that the quantities
$\rho_k(t)$ are, in finite networks, of stochastic nature, is
neglected.  Instead, deterministic rate equations are considered,
taking into account the changes in time of the partial densities, due
to the different steps that the evolution of the model can take.

In the case of the CP, the quantities $\rho_k(t)$, given by
\begin{equation}
  \rho_k(t) = \frac{n_k(t)}{N P(k)},
\end{equation}
where $n_k(t)$ is the number of particles in vertices of degree $k$,
can be interpreted equivalently as the relative densities of particles
in vertices of degree $k$, or the probabilities that a given vertex of
degree $k$ contains a particle. In a step of the CP dynamical
evolution, the partial density $\rho_k(t)$ can decrease due to the
annihilation of a particle in a vertex $k$ (with rate $1$), or can
increase by the generation of an offspring in a vertex $k'$, nearest
neighbor of $k$ (with rate $\lambda/ k'$).  Therefore, the rate
equations for the partial densities in a network characterized by a
degree distribution $P(k)$ and degree correlations given by the
conditional probability $P(k'|k)$ can be written as
\cite{Castellano:2006}
\begin{equation}
  \frac{\partial \rho_k(t)}{\partial t} = - \rho_k(t) +  \lambda  k
  [1-\rho_k(t)] \sum_{k'} \frac{P(k' | 
    k) \rho_{k'} (t)}{k'}.
  \label{eq:1}
\end{equation}
Given Eq.~(\ref{eq:1}), $\rho_k=0$ is always
a solution.  The conditions for the presence of non-zero steady states
can be obtained by performing a linear stability analysis~\cite{Boguna:2002}.
Neglecting higher order terms, Eq.~(\ref{eq:1}) becomes
\begin{equation}
  \frac{\partial \rho_k(t)}{\partial t} \simeq \sum_{k'} L_{k k'}
  \rho_{k'}(t) \equiv \sum_{k'} \left(  -\delta_{k, k'} + \lambda  k
    \frac{P(k'|k)}{k'} \right)\rho_{k'}(t).
\end{equation}
It is easy to see that the Jacobian matrix $L_{k k'}$ has a unique
eigenvector $v_k =k$ and a unique eigenvalue $\Lambda =
\lambda-1$. Therefore, a nonzero steady state is only possible for
$\Lambda>0$, which translates in a critical threshold for the
absorbing state phase transition
\begin{equation}
  \lambda_c=1,
\end{equation}
independent of the degree distribution and the correlation pattern.

To get more detailed information on the process, and in
particular on the shape of the order parameter as a function of the rate
$\lambda$, we restrict our attention to uncorrelated networks.  In this
case, Eq.~(\ref{eq:1}) reads
\begin{equation}
  \frac{\partial \rho_k(t)}{\partial t} = - \rho_k(t) +
  \lambda \frac{k}{\avk} [1-\rho_k(t)] \rho(t).
  \label{eq:2}
\end{equation}
Imposing the steady state condition, $\partial_t \rho_k(t)=0$, yields
the nonzero solutions
\begin{equation}
  \rho_k = \frac{\lambda k \rho / \avk}{1+\lambda k \rho / \avk} 
  \label{eq:4}
\end{equation}
where $\rho_k$ is now independent of time. 
By combining Eq.~(\ref{eq:4}) with the
definition of $\rho$, one obtains the self-consistent equation for
the order parameter $\rho$,
\begin{equation}
  \rho =  \frac{\lambda \rho }{\avk} \sum_k  \frac{k
    P(k)}{1+\lambda k \rho / \avk},
  \label{eq:6}
\end{equation}
that depends on the full degree distribution.

In the case of
SF networks, for which the degree distribution in the continuous
degree approximation is given by
$P(k) = (\gamma-1) m^{\gamma-1} k^{-\gamma}$, with
$m$ the minimum degree in the network, the solution will depend on the
degree exponent $\gamma$. Substituting the summation by an integral in
Eq.~(\ref{eq:6}), we obtain in the \textit{infinite network size
  limit} (i.e.  when the degree belongs to the range $[m, \infty]$) the
expression 
\begin{equation}
  \rho = F\left[1,\gamma-1,\gamma,-\frac{\avk}{\lambda \rho m}\right],
\end{equation}
where $F[a,b,c, z]$ is the Gauss hypergeometric
function~\cite{abramovitz}.  To evaluate the critical behavior for
small $\rho$, we invert this expression using the asymptotic expansion
of the hypergeometric function for low densities \cite{abramovitz},
obtaining the result
$\rho(\lambda) \sim (\lambda-1)^\beta$, with $\beta=1/(\gamma-2)$ for
$2<\gamma<3$ and $\beta=1$ for $\gamma>3$, presenting additional
logarithmic corrections at $\gamma=3$. 

Right at the critical point, $\lambda=1$, the particle density is
expected to decay as a power law of time, $\rho(t) \sim t^{-\theta}$
\cite{Marro}, defining a new, temporal, critical exponent. This
exponent can be estimated within HMF, by considering the time
evolution of the total density at $\lambda=1$, namely,
\begin{equation}
  \frac{ \partial \rho(t)}{ \partial t} = \sum_k P(k) \frac{ \partial
    \rho_k(t)}{ \partial t} = -\frac{\rho(t)}{\avk} \sum_k k \rho_k(t).
  \label{eq:22}
\end{equation}
To close this equation, we use a quasi-static
approximation~\cite{Catanzaro:2005b}, which can be justified in terms
of an adiabatic approximation for the full Langevin theory for the CP
(see Sec.~\ref{sec:cp-dynam-uncorr}). In essence, we consider that,
even at the critical point, where no steady-state is present, the
partial densities relax to a quasi-stationary state, where they take
the form given by Eq.~(\ref{eq:4}). In this case, for SF
networks in the continuous degree approximation, Eq.~(\ref{eq:22})
will read
\begin{equation}
  \frac{ \partial \rho(t)}{ \partial t} \simeq 
  -\rho(t)   F\left[1,\gamma-2,\gamma-1,-\frac{\avk}{\rho(t) m}\right], 
\end{equation}
Using the asymptotic approximation
for the hyper-geometric function, valid for low density, we obtain a
decay exponent in infinite networks given by $\rho(t)\sim t^{-\theta}$,
with $\theta=\beta$ for all $\gamma$
(logarithmic corrections being again present at $\gamma=3$).

\section{Finite-size scaling for the CP in complex networks}
\label{sec:finite-size-scaling}

The exponents obtained within HMF theory in the previous Section
correspond to the thermodynamic limit of a SF network of infinite
size. Checking their accuracy in numerical simulations becomes thus a
nontrivial task, particularly close to the critical point, due to the
effects of finite network sizes.  Indeed, because of the small-world
property, the number of neighbors that can be reached starting from a
certain node grows exponentially or faster with the geodesic
distance. This implies that, even for large networks, just a few steps
are sufficient to probe the finiteness of the system. Moreover, in SF
networks, local topological properties show very strong fluctuations,
increasing with the size of the network.

For general critical phenomena, the theory of finite-size scaling
(FSS) \cite{privman90} has successfully overcome this problem for
processes taking place on regular lattices, allowing the detection of
the signature of continuous phase transitions even in very small
systems.  For absorbing state phase transitions, FSS is based on
the observation that, even below the critical point, the density of
active sites in surviving runs $\rho_s$ reaches a quasi-steady state
whose average is a decreasing function of the system size, and that
can be expressed as a homogeneous scaling function of both the system
size and the distance to the critical point. In the case of networks,
system size is replaced by the number of vertices, and the surviving
density is assumed to fulfill the relation
\cite{zanette01:_critic,PhysRevE.67.026104}
\begin{equation}
    \rho_{s}(\Delta, N)=N^{-\beta/\bar{\nu}}f(\Delta N^{1/\bar{\nu}}),
  \label{eq:14}
\end{equation}
where $f(x)$ is a scaling function that behaves as $f(x)\sim
x^{\beta}$ for $x\rightarrow \infty$ and $f(x)\sim \mbox{const}$ for
$x\rightarrow 0$. 

Mean-field theory for homogeneous networks predicts the exponents
$\beta=1$ and $\bar{\nu}=2$. For the case of SF, Ref.~\cite{Hong:2007}
proposed a phenomenological Langevin equation for the particle
density, taking the form
\begin{equation}
  \frac{ d \rho(t)}{d t}= \Delta \rho(t) - b \rho(t)^2 -d
  \rho(t)^{\gamma-1} + \sqrt{\rho(t)} \eta(t),
\label{KoreanLE}
\end{equation}
where $\eta(t)$ is an uncorrelated Gaussian noise. Assuming a scaling
form for the surviving density given by Eq.~(\ref{eq:14}), and by
means of a droplet-excitation argument, the authors of~\cite{Hong:2007}
found that $\beta=1/(\gamma-2)$ and $\bar{\nu}=(\gamma-1)/(\gamma-2)$
for $\gamma<3$, independent of the network cutoff, whenever $k_c(N) >
N^{1/\gamma}$ \cite{hong07:_comment}.

In Ref.~\cite{Castellano:2008}, this issue  was pursued by focusing on
the FSS form of survival probability, which at the critical point, and
in networks of size $N$, was assumed to be 
\begin{equation}
  \label{survival:1}
  S(t,N)=t^{-\delta}f(t/t_c(N)).
\end{equation}
The scaling function $f(x)$ is constant for small values of the
argument and cutoff exponentially for $x \gg 1$. $t_c(N)$ is a
characteristic cutoff time that, according to standard mean-field FSS
theory should scale as $t_c(N) \sim N^{1/2}$ for homogeneous
networks~\cite{Marro}. By means of a mapping to a biased,
one-dimensional random walk, the authors of~\cite{Castellano:2008}
found $\delta=1$, while the characteristic time showed the form, for
heterogeneous networks, $t_c(N) \sim \sqrt{N/g}$,
where $g$ is defined in Eq.~(\ref{eq:10}), and is thus dependent on
the degree cutoff. This surprising result, well confirmed
by numerical simulations \cite{Castellano:2008}, is in
strong disagreement with results of
Refs.~\cite{hong07:_comment,Hong:2007}, in which no cutoff dependence
was claimed.

In order to fully ascertain the correct FSS behavior of CP in SF
networks, we go beyond mean-field and phenomenological theories
and tackle the full problem, taking into account its implicit
stochastic fluctuations (particularly important in the vicinity of a
critical point) by means of a Langevin approach. This problem is
considered in the next Section.

\section{Langevin approach for the CP on networks}
\label{sec:langevin-approach-cp}

\subsection{Generic formalism}

To account for the stochastic fluctuations of the CP close to
the critical point, we derive here a Langevin equation describing the
concentration $\rho_k(t)$ or, alternatively, the number of active
sites of degree $k$, $n_k(t)$. Our derivation follows closely the
method developed in~\cite{Catanzaro:2005b}.
We start by deriving exact equations for the microscopic dynamics (at
the vertex level) of the process. Let $\sigma_i(t)$ be a random binary
variable taking value $\sigma_i(t)=1$ if node $i$ is occupied by a
particle at time $t$ and $\sigma_i(t)=0$ otherwise. Thus, the state of
the process at time $t$ is completely determined by the state vector
${\bf \Sigma}(t)=\{\sigma_1(t),
\sigma_2(t),\cdots,\sigma_N(t)\}$. Variables $\sigma_i(t)$ can undergo
only two types of transition events:
\begin{enumerate}
\item $\sigma_i(t)=1 \to \sigma_i(t+dt)=0$: Vertex $i$ was occupied
  by a particle at time $t$, and the particle annihilated during the
  time interval $[t,t+dt]$.
\item $\sigma_i(t)=0 \to \sigma_i(t+dt)=1$: Vertex $i$ was empty at
  time $t$ and it received an offspring from an occupied nearest
  neighbor during the time interval $[t,t+dt]$.
\end{enumerate}
Assuming that the temporal occurrence of these events follows Poisson
processes, the previous two events can be encoded into a single
dynamical equation that describes the evolution of $\sigma_i(t)$ after an
increment of time $dt$ as
\begin{equation} 
  \label{micro}
  \sigma_i(t+dt)=\sigma_i(t)\zeta_i(dt)+[1-\sigma_i(t)]\eta_i(dt),
\end{equation}
where $\zeta_i(dt)$ and $\eta_i(dt)$ are dichotomous random variables
taking values
\begin{equation}
  \label{xi}
  \zeta_i(dt)=\left\{
    \begin{array}{ll}
      0 & \mbox{with probability  } dt\\[0.5cm]
      1 & \mbox{with probability  } 1-dt
    \end{array}
  \right.
\end{equation}
and
\begin{equation}
  \label{eta}
  \eta_i(dt)=\left\{
    \begin{array}{ll}
      1 & \mbox{with probability  } \lambda dt \displaystyle{\sum_j
        a_{ij} \sigma_j(t)\frac{1}{k_j}}\\[0.5cm] 
      0 & \mbox{with probability  } 1-\lambda dt \displaystyle{\sum_j
        a_{ij} \sigma_j(t)\frac{1}{k_j}} 
    \end{array}
  \right. .
\end{equation}
Eqs.~(\ref{xi}) and~(\ref{micro}) describe the annihilation of
particles, while Eq.~(\ref{eta})  corresponds to the creation from
occupied nearest neighbors.

The set of random variables $\{\zeta_i(dt) ; i=1,\cdots,N\}$ are
statistically independent of each other and of the conjugate
random variables $\{\eta_i(dt) ; i=1,\cdots,N\}$. On the other hand,
variables $\{\eta_i(dt) ; i=1,\cdots,N\}$ are not totally independent
since they may involve common events inducing correlations among
them. For example, imagine two empty vertices, A and B, each of degree
$1$, connected to the same occupied vertex C. Because of the CP
dynamics, during a particle reproduction event at vertex C, the
particle must choose only one of its neighbors to send the
offspring. Therefore, if vertex A gets the offspring, vertex B cannot
receive it and vice-versa, inducing thus correlations between the
random variables $\eta_A(dt)$ and $\eta_B(dt)$. However, it is easy to
see that these correlations are of order $dt^2$ and can be then safely
neglected. In any case, this effect only exists in networks with a
quenched topology. In contrast, the annealed
network topology changes faster than the CP dynamics and, therefore,
such correlations are absent.

Equations~(\ref{micro}), (\ref{xi}), and (\ref{eta}) describe the
evolution of the state of the system at the most detailed possible level of
description by specifying the precise state of each and every one of
the vertices of the network. This description, although exact, is not
very useful to derive general properties of the system, which are
better described by coarse-grained quantities. In heterogeneous random
networks with given degree distribution $P(k)$ and degree-degree
correlations $P(k'|k)$, the degree of vertices $k$ is the most
appropriate indicator of the different classes of vertices. Therefore,
we consider all vertices with the same degree to be statistically
equivalent. Following these ideas, let $n_k(t)$ be the number of
active vertices of degree $k$ at time $t$, that is,
\begin{equation}
n_k(t) \equiv \sum_{i \in k} \sigma_i(t).
\end{equation}
As we can see, $n_k(t)$ is the sum of a large number of random
variables that are nearly statistically independent in the quenched
version of the network and totally independent in its annealed
version. Therefore, by invoking the central limit theorem, we expect
this variable to follow a Gaussian distribution and, consequently, to
follow a Langevin dynamics. To derive the specific form of this
Langevin equation, we need to calculate the infinitesimal moments of
$n_k(t)$, which can be done using Eqs.~(\ref{micro}), (\ref{xi}), and
(\ref{eta}). Using the results in
Appendix~\ref{sec:appendix-calc-coarse-grain}, we can finally write
the corresponding Langevin equation for the CP on annealed networks, namely 
\begin{eqnarray}
  \label{Langevin:1}
  \frac{d n_k(t)}{dt} &=&  -n_k(t)+\lambda \left[1-\rho_k(t)\right]
  \sum_{k'}P(k|k')n_{k'}(t) \\ \nonumber 
 &+ & \xi_k(t) \sqrt{n_k(t)+\lambda
    \left[1-\rho_k(t)\right] \sum_{k'}P(k|k')n_{k'}(t)} , 
\end{eqnarray}
where $\rho_k(t)=n_k(t)/P(k)N$ is the relative density of active
vertices of degree $k$ and $\{ \xi_k(t), k=1,\cdots,k_c\}$ are
Gaussian white noises (with zero mean and unit variance)
uncorrelated among them. Eq.~(\ref{Langevin:1}) implicitly assumes
that $n_k(t)$ is a continuous variable. This approximation is
reasonable as long as $n_k(t)\gg 1$, which is usually the case in very
large systems, when we consider steady-state properties.

Equation~(\ref{Langevin:1}) is one of the main results of this paper
and is also the starting point for our subsequent analysis. As one
immediately recognizes, the drift term in Eq.~(\ref{Langevin:1})
corresponds to the standard mean-field approximation derived in
Eq.~(\ref{eq:1}).  It is easy to see that the potential associated
with this drift term has a stable minimum whenever
$\lambda>\lambda_c=1$ which does not depend on the particular
correlation pattern given by $P(k|k')$. The position of this minimum
corresponds to the steady solution in the active phase in the
thermodynamic limit. The diffusion term, on the other hand, points to
a process with multiplicative noise that, as we shall see, has
important implications when the system is close to its critical point
in finite size systems.

\subsection{Uncorrelated random networks} 
\label{sec:cp-dynam-uncorr}

Finding solutions of Eq.~(\ref{Langevin:1}) for networks with general
degree-degree correlations is a rather difficult task. In this paper,
we focus on the simplest (but instructive) case of uncorrelated random
networks with a given degree distribution $P(k)$. For this class of
networks, the transition probability takes the simple form
$P(k|k')=kP(k)/\langle k \rangle$ which allows us to write
Eq.~(\ref{Langevin:1}) as
\begin{eqnarray}
  \label{Langevin:1_uncorr}
  \frac{d \rho_k(t)}{dt} &=&-\rho_k(t)+\lambda \frac{k}{\langle k
    \rangle}\left[1-\rho_k(t)\right] \rho(t) \\ \nonumber
 & +&\sqrt{\frac{1}{N_{k}}\left( \rho_k(t)+\lambda \frac{k}{\langle k
        \rangle}\left[1-\rho_k(t)\right] \rho(t)\right)} \xi_k(t), 
\end{eqnarray}
where $\rho(t)=\sum_k n_k(t)/N=\sum_{k}P(k) \rho_{k}(t)$ is the global
concentration of active nodes at time $t$ and we have divided
Eq.~(\ref{Langevin:1}) by the number of vertices of degree $k$,
$N_{k}=NP(k)$.  Analogously, we can write a Langevin equation for
$\rho(t)$ as
\begin{equation}
  \label{Langevin:2}
  \frac{d \rho(t)}{dt} = \rho(t)\left(\Delta-\lambda \sum_k
    \frac{kP(k)}{\langle k \rangle}\rho_k(t)\right) 
\end{equation}
\[
  +\sum_k P(k)\sqrt{\frac{1}{N_{k}}\left( \rho_k(t)+\lambda \frac{k}{\langle k
        \rangle}\left[1-\rho_k(t)\right] \rho(t)\right)} \xi_k(t), 
\]
where we have defined $\Delta \equiv \lambda-1$ so that the critical
point corresponds to $\Delta=0$. 

Eq.~(\ref{Langevin:2}) is not yet a closed equation for $\rho(t)$
because both the drift and diffusion terms involve the partial
densities $\rho_k(t)$.  To close it, we use an adiabatic
approximation~\cite{Gardiner:2004}. From Eq.~(\ref{Langevin:2}) we
know that close to the critical point, $\Delta\approx0$, $\rho(t)$ is
a slowly varying variable. This is due to the fact that the first term
in the right hand side of Eq.~(\ref{Langevin:2}) is of order higher
than $\rho$. On the other hand, $\rho_{k}(t)$ is a variable that
relaxes exponentially fast to its quasi-equilibrium state since the
lowest order in Eq.~(\ref{Langevin:1_uncorr}) is linear in
$\rho_{k}$~\footnote{It is worth mentioning that this separation of
  time scales between the partial quantities $\rho_k$ and the global
  one $\rho$ has also been observed in other dynamics like the
  $A+A\longrightarrow \emptyset$ diffusion-annihilation
  process~\cite{Catanzaro:2005b} or the voter
  models~\cite{Sood:2005}.}.  The adiabatic approximation consists in
neglecting the term $d\rho_{k}(t)/dt$ in front of $\rho_{k}(t)$ and
assuming that $\rho_{k}$ is a stochastic variable that evolves much
faster than $\rho(t)$. Thus, setting $d\rho_{k}(t)/dt=0$ in
Eq.~(\ref{Langevin:1_uncorr}) and solving for $\rho_k(t)$, we obtain
\begin{equation}
  \label{adiabatic}
  \rho_k(t)  \approx \frac{\lambda k \rho(t)}{\langle k
    \rangle+\lambda k \rho(t)} 
\end{equation}
\[
    + \frac{\langle k \rangle}{\langle k \rangle+\lambda k
      \rho(t)}\sqrt{\frac{1}{N_{k}}\left( \rho_k(t)+\lambda
      \frac{k}{\langle k \rangle}\left[1-\rho_k(t)\right]
      \rho(t)\right)} \xi_k(t).
\]
The noise term in this equation is subdominant due to its dependence
on the size of the system. Thus, replacing the dominant term in the
diffusion one, we finally obtain
\begin{equation}
  \label{adiabatic2}
  \rho_k(t) \approx \frac{\lambda k \rho(t)}{\langle k \rangle+\lambda
    k \rho(t)} + \sqrt{\frac{1}{N_{k}} \frac{2 \lambda k
      \rho(t)\langle k \rangle^2}{[\langle k \rangle+\lambda k
        \rho(t)]^3}} \xi_k(t).
\end{equation}
In this way, we obtain an expression for the partial densities
$\rho_k$ as a function of $k$ and $\rho(t)$ only. Replacing this
expression in Eq. (\ref{Langevin:2}) and keeping only the first order
in $N_k^{-1}$, we obtain
\begin{eqnarray}
  \frac{d \rho(t)}{dt} & = & \rho(t)\left(\Delta-\lambda \sum_k
    \frac{kP(k)}{\langle k \rangle}\frac{\lambda k \rho(t)}{\langle k
    \rangle+\lambda k \rho(t)}\right)\\ \nonumber
  &+&\sum_k P(k) \sqrt{\frac{1}{N_{k}} \frac{2 \lambda k \rho(t)\langle k
    \rangle^2}{[\langle k \rangle+\lambda k \rho(t)]^3}} \xi_k(t). 
\end{eqnarray}
Notice that the sum of statistically independent Gaussian white noises
is another Gaussian white noise whose variance is the sum of the
individual variances. Thus, the diffusion term in the last equation
is, indeed, a Gaussian white noise. Therefore, we can finally write
\begin{equation}
  \label{Langevin:3}
  \frac{d \rho(t)}{dt} =\rho(t)\left(\Delta-\lambda
  \Theta[\rho(t)]\right)+\sqrt{\frac{2\lambda \rho(t)}{N}
    \Lambda[\rho(t)]} \xi(t),
\end{equation}
where
\begin{equation}
\Theta[\rho(t)]\equiv \sum_k \frac{kP(k)}{\langle k
  \rangle}\frac{\lambda k \rho(t)}{\langle k \rangle+\lambda k
  \rho(t)} 
\label{eq:7}
\end{equation}
and 
\begin{equation}
\Lambda[\rho(t)]\equiv \sum_k \frac{kP(k)}{\langle k
  \rangle}\frac{\langle k \rangle^3}{[\langle k \rangle+\lambda k
  \rho(t)]^3}. 
\label{eq:7bis}
\end{equation}

Eq.~(\ref{Langevin:3}) is now a closed equation for the total density
of active vertices $\rho$ which must be solved with an absorbing
boundary at $\rho=0$ and a reflecting one at $\rho=1$. As we can see
from Eq.~(\ref{Langevin:3}), there is an explicit dependence on the
size of the network $N$ in the diffusion term of the Langevin
equation. This size dependence, together with the specific functional
forms of $\Theta[\rho]$ and $\Lambda[\rho]$ will determine the finite
size behavior of the system near the critical point.

\section{CP in annealed scale-free networks}
\label{sec:CPSF}
In this section we focus on heterogeneous networks with a power law
degree distribution $P(k) \sim k^{-\gamma}$ with $k \in [m,M]$, where
$2< \gamma <3$, and
$M = N^{1/\omega}$ is the degree upper cutoff.  
In particular, we consider the case $\omega \ge \gamma-1$, so that
the average maximum of the degree distribution
$k_c \sim M = N^{1/\omega}$ is a hard cutoff
(Sec.~\ref{sec:annealed-scale-free}).
The case $\omega < \gamma-1$ will be considered in
Sec.~\ref{sec:effect-outliers}.

Given the  form of the degree distribution it is possible to
evaluate explicitly the functional form of $\Theta[\rho]$, that
determines the dynamical properties of the CP.  From its
definition, Eq.~(\ref{eq:7}), it is easy to see that in the limit of
small density $\Theta[\rho]$ has two different functional forms
depending on whether $\rho$ is larger or smaller than the quantity
$\avk/\lambda k_c$, $k_c$ being the network cutoff. Thus we have
(see Appendix~\ref{sec:calculation-thetarho}):
\begin{equation}
  \label{theta}
  \Theta[\rho]=\left\{
    \begin{array}{lcr}
      g \lambda \rho
      & \rho \ll \frac{\langle k
        \rangle}{\lambda k_c} & \mbox{region II} \\[0.5cm] 
      C(\gamma) \left( \frac{\lambda \rho}{\langle k
          \rangle}\right)^{\gamma-2} &  
      \frac{\langle k \rangle}{\lambda k_c} \ll \rho \ll 1 &\mbox{region I}
    \end{array}
  \right. ,
\end{equation}
where $g=\langle k^2 \rangle/\langle k \rangle^2$ and
\begin{equation}
C(\gamma) =   m^{\gamma-2} \Gamma(\gamma-1) \Gamma(3-\gamma).
  \label{eq:9}
\end{equation}
We denote the regime for small $\rho$ as region II, and the regime for
larger $\rho$ as region I.
Analogously, we can evaluate the behavior of the function $\Lambda$ as
\begin{equation}
  \label{lambda}
  \Lambda[\rho]=\left\{
    \begin{array}{lr}
      1      & \mbox{region II} \\[0.5cm] 
      1-\widetilde{C}(\gamma) \left( \frac{\lambda \rho}{\langle k
          \rangle}\right)^{\gamma-2} &  
      \mbox{region I}
    \end{array}
  \right. ,
\end{equation}
with
\begin{equation}
\widetilde{C}(\gamma)=\frac{1}{2}m^{\gamma-2}(\gamma-2)(\gamma-1) \Gamma(1+\gamma)\Gamma(1-\gamma).
\end{equation}
As we can see, the correction term in the region I is always very
small as compared to $1$. Therefore, in the rest of the paper we
consider that $\Lambda[\rho]=1$.

\begin{figure}
\begin{center}
\includegraphics[width=\columnwidth]{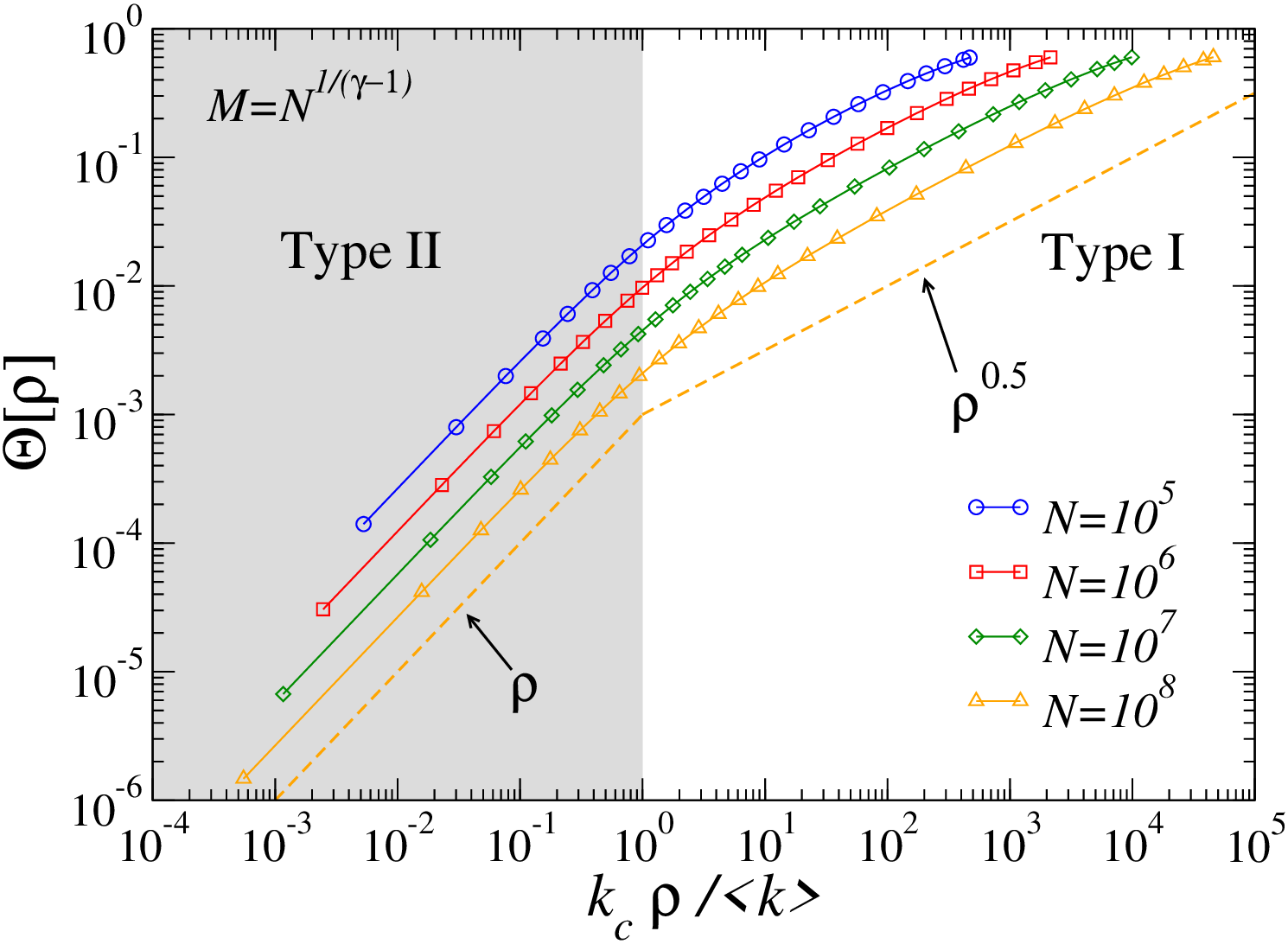}
\end{center}
\caption{Numerical evaluation of the function $\Theta[\rho]$ as a
  function of $k_c\rho/\langle k \rangle$ for different network
  sizes. The degree exponent is $\gamma=2.5$ and we use
  $\omega=\gamma-1$. In the type II region, it is clearly visible a
  linear behavior, in agreement with Eq.~(\ref{theta}).  In the type I
  region, convergence towards the theoretical expression given by
  Eq.~(\ref{theta}) is much slower. }
\label{drift}
\end{figure}
In Fig.~\ref{drift} the behavior of $\Theta[\rho]$ is evaluated
by numerically performing the summation in Eq.~(\ref{eq:7}).
The linear behavior
for small densities is very well obeyed. Instead, the scaling of the
region I is not cleanly observed even for the largest network
considered ($N=10^8$).  This is due to the fact that region I is
surrounded by two slow crossovers, one for $\rho \approx \avk/k_c$
(where the transition between region I and II takes place) and the
other for $\rho=1$ (where $\Theta$ becomes independent of
$\rho$). This has the consequence that some of the theoretical
predictions made using the simple approximation given in Eq.~(\ref{theta})
are difficult to observe except for extremely large system sizes.

\begin{figure}
\begin{center}
\includegraphics[width=\columnwidth]{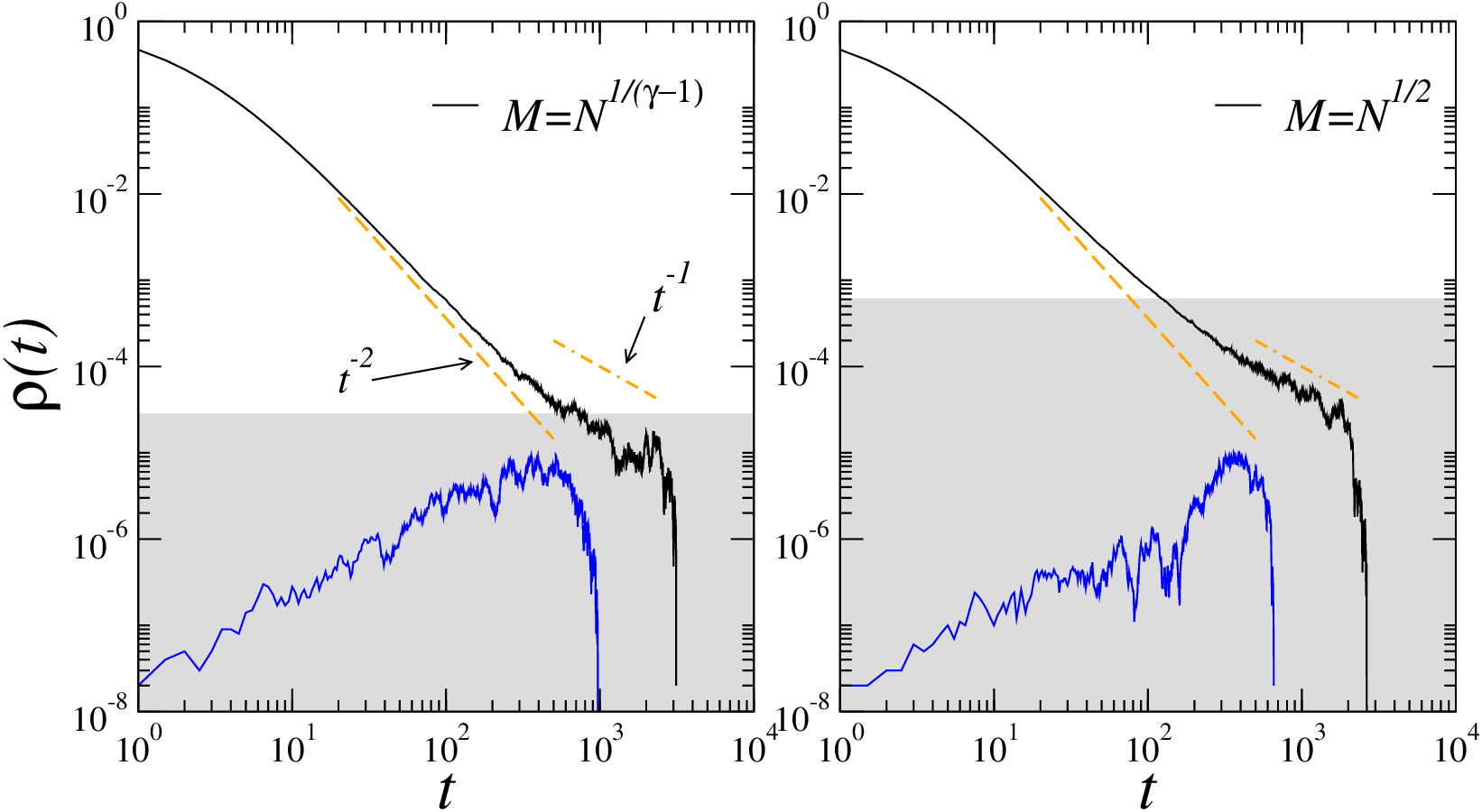}
\end{center}
\caption{
  Example of two different types of experiment to study
  the CP dynamics at criticality performed in an annealed network of
  size $N=10^8$, $\gamma=2.5$, and $m=2$. 
  The top (black) curve is the density decay starting from a fully
  active network. The bottom (blue) curve corresponds to a spreading
  experiment starting from a single active vertex. 
  The left plot corresponds to
  the hard cutoff $M=N^{1/(\gamma-1)}$ and the right plot to
  $M=N^{1/2}$. Grey areas depict the values of
  $\rho$ in the domain $\rho \in [N^{-1},\langle k \rangle
  k_{c}^{-1}]$. In all cases trajectories are for a single run in an
  instance network.}
\label{experiments}
\end{figure}

As a consequence of the form of $\Theta[\rho]$,
the behavior of the system at criticality strongly depends
on the type of experiment performed to probe the absorbing
transition, see Fig.~\ref{experiments}. 
Indeed, experiments with stochastic trajectories exploring
the region $\rho \gg \langle k \rangle/\lambda k_c$ feel a drift of
the form
\begin{equation}
  \Psi[\rho]\simeq\rho\left[ \Delta-C(\gamma) \left( \frac{\lambda
        \rho}{\langle k \rangle}\right)^{\gamma-2}\right]
  \mbox{\hspace{0.1cm} type I drift}. 
\label{driftII}
\end{equation}
Instead,  any experiment such that trajectories mainly stay
in the region 
$\rho \ll \langle k \rangle/\lambda k_c$ feels a drift
term of the form
\begin{equation}
  \Psi[\rho]=\rho(\Delta -\lambda^2 g \rho) \mbox{ \hspace{1cm} type
    II drift}.
\label{driftI}
\end{equation}

The quantity $g$, that will play a fundamental role in the rest of the paper,
diverges with the cutoff $k_c$ as $k_c^{\gamma-3}$ for $\gamma<3$ 
(and $\omega>\gamma-1$).
Notice that for $\gamma>3$ the leading order is linear in both cases.
It is also worth stressing that if one lets $k_c$ diverge, regime II
disappears and one is left only with regime I, that coincides with
what is found using HMF on infinite networks
(Sec.~\ref{sec:heter-mean-field}).  However, in any finite network,
when the density gets small it is the drift of type II that rules the
dynamics.

Based on the explicit expression of $\Theta[\rho]$
we now provide a qualitative and quantitative description
of the three types of experiment that explore the critical
properties of the CP dynamics: A) density decay at criticality,
B) spreading experiments and C) surviving runs.

\subsection{Density decay at criticality} 

Starting from a configuration full of active vertices at $t=0$, the
concentration of active vertices is monitored as a function of time
until the trajectory is trapped at the absorbing boundary. Then an
average is performed over a large number of different runs up to a
time such that all runs have survived. In this case, after a initial
timescale $t_\times$, the system first feels the type I drift and
after a crossover time $t^*$, at very low concentrations, the type II
one.  Inserting Eq.~(\ref{driftII}) into Eq.~(\ref{Langevin:3}), a
pure drift of the type I predicts a behavior $\rho_{I}(t) \sim
t^{-\theta}$ with $\theta=1/(\gamma-2)$.  Inserting Eq.~(\ref{driftI})
gives instead $\rho_{II}(t)\sim (gt)^{-1}$.  The crossover between the
two types of behavior occurs for a time $t^*$ such that
$\rho_{II}(t^*) \simeq \avk/k_c$, i.e.  $t^* \sim k_c/(g \avk) \sim
\avk k_c^{\gamma-2}$. A third time scale defines the survival time of
the different runs that, as we will see in the next subsection, scales
as $t_c \sim \sqrt{N/g}$.

Fig.~\ref{experiments} shows simulation results for this type of
experiment (top curves) in annealed networks
with $\gamma=2.5$, $m=2$, for $\omega=2$ or $\omega=\gamma-1$
for a single run starting from a fully active network. Different
colors (grey and white) indicate the different regions depending on
the shape of the drift term. The first thing to notice is that, in the
case of $\omega=\gamma-1$, the region that corresponds to the type
I drift is wider as compared to the case $\omega=2$.
Nevertheless, even in this optimal case, we do not observe
cleanly the two different values of the exponent $\theta$.
For comparison purposes, in Fig.~\ref{experiments} we also plot
functions $t^{-2}$ and $t^{-1}$
that would correspond to the pure type I and II behaviors for
$\gamma=2.5$. The exponent $\theta$ approaches but does not reach the
theoretical value $\theta=2$ even though simulations are performed in
networks of size $N=10^8$. The situation in the case of $\omega=2$
is even worse because the crossover happens at shorter times
and the value $\theta=2$ is even more difficult to observe.
\begin{figure}
\begin{center}
\includegraphics[width=\columnwidth]{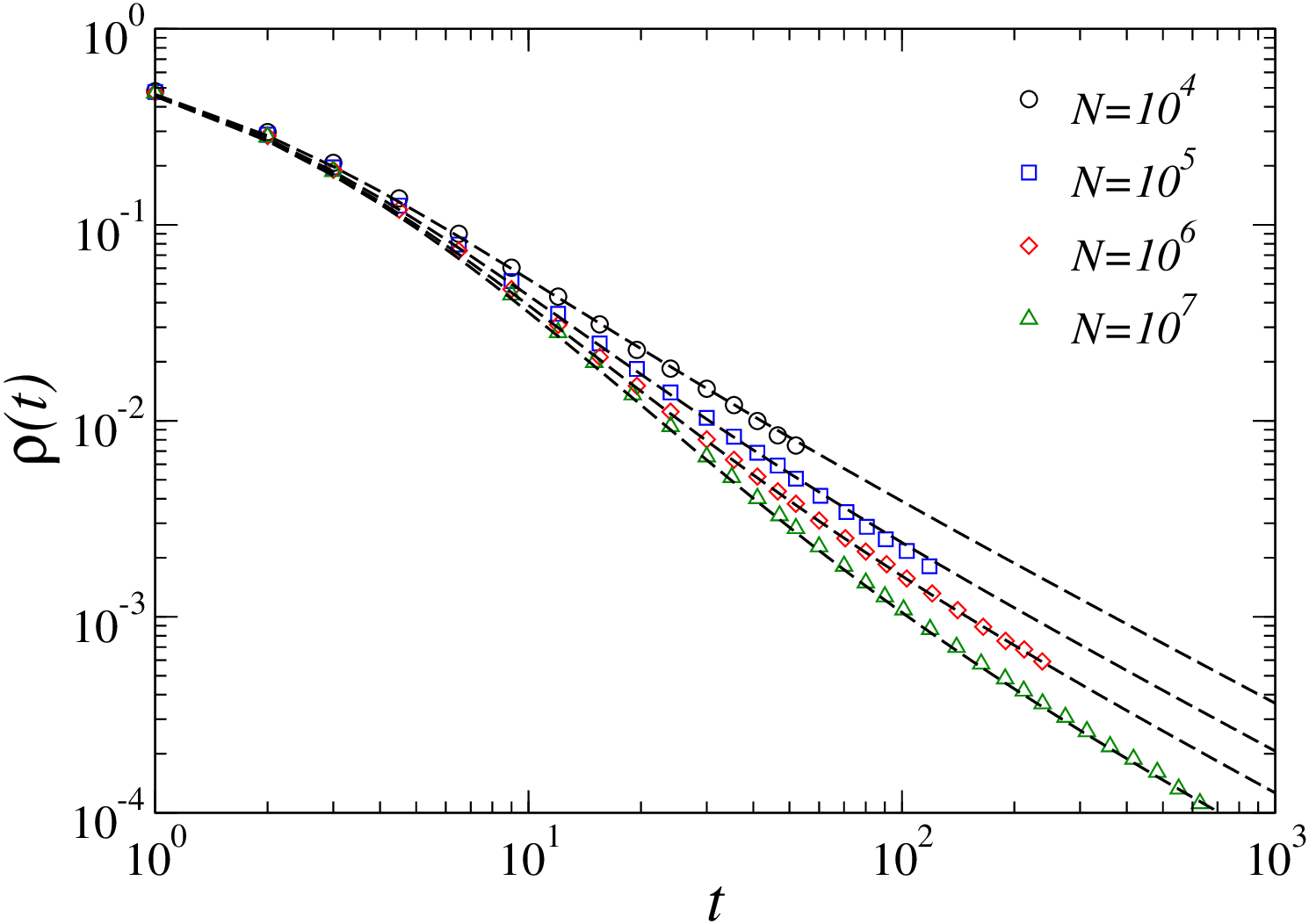}
\end{center}
\caption{Evolution of the concentration $\rho(t)$ starting from a
  fully infected network for different network sizes, $\gamma=2.5$,
  and $\omega=2$. Results are averaged over $100$ realizations in a
  single instance network. Long dashed lines correspond to the
  numerical solution of the set of Eqs.~(\ref{eq:2}), using as input
  the empirical degree sequence used in the simulations. The very nice
  agreement between both sets of curves justifies our approximation.}
\label{rho_fully_infected}
\end{figure}

\begin{figure}
\begin{center}
\includegraphics[width=\columnwidth]{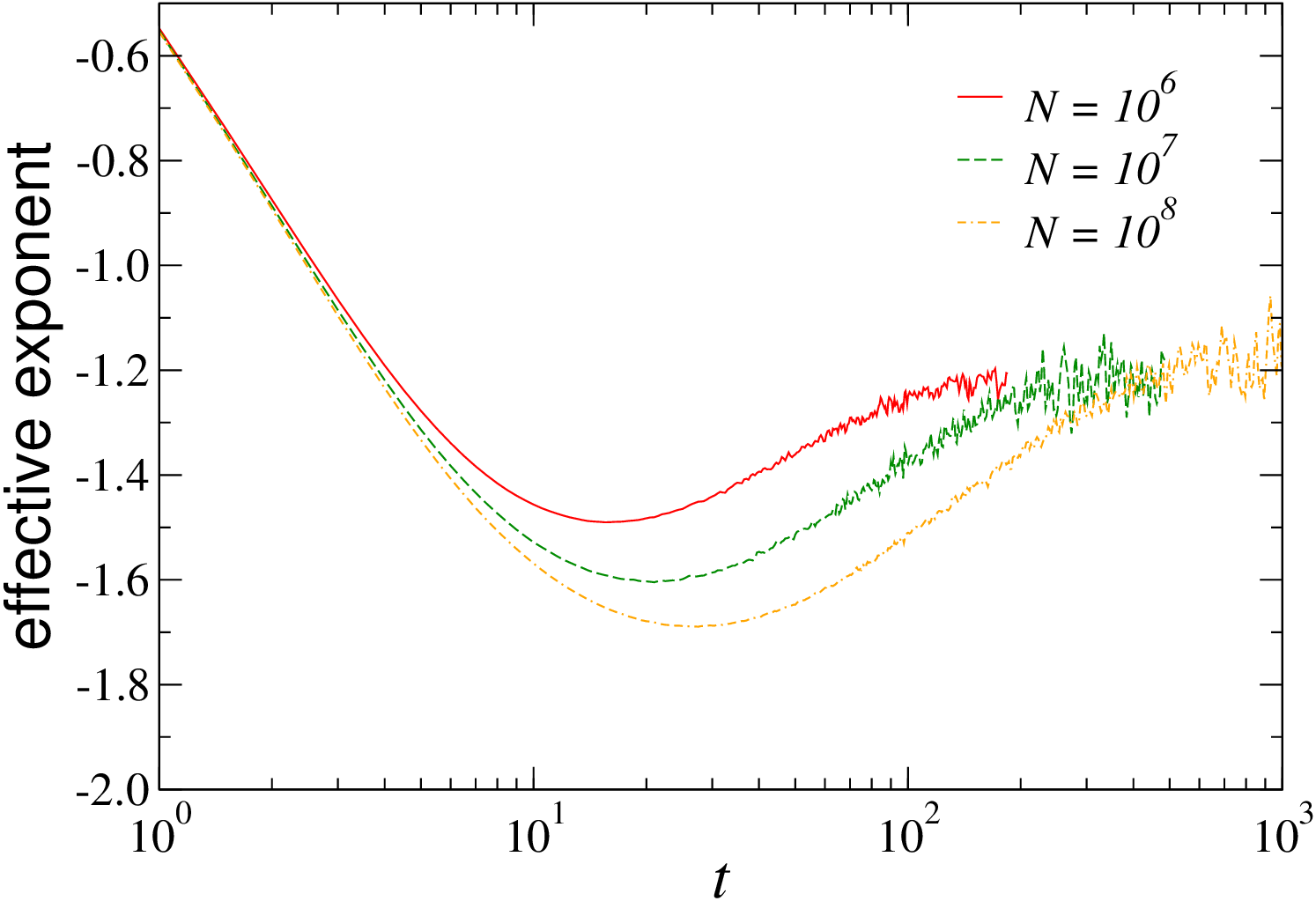}
\end{center}
\caption{Local effective exponent
as a function of time for $\gamma=2.5$, $\omega=2$ and various values of
$N$. The exponent corresponding to the discrete time $t_i$ is given
by the slope of the line joining $\rho(t_{i-1})$ and $\rho(t_{i+1})$
in log-log scale.}
\label{effectiveexp}
\end{figure}

The same is observed in Fig.~\ref{rho_fully_infected}, where we show
the same as Fig.~\ref{experiments} for networks with $\gamma=2.5$,
$\omega=2$ and different network sizes but averaging over $100$ runs
of the process over the same instance network.
Additional information is provided by the local effective
exponent of the temporal decay as a function of time
(Fig.~\ref{effectiveexp}).
The effective exponent decreases initially quite fast, 
but even for $N=10^8$ the crossover to regime II takes place well
before the asymptotic value $\theta=1/(\gamma-2)=2$ is reached.
Eventually the effective exponent sets to a constant value, that is,
quite surprisingly, close to 1.2 instead of the expected value $\theta=1$.

The reason why we do not see convincing numerical evidence of any
of the two scaling exponents expected from the theory
is that the separation of time scales between
$t_{\times}$, $t^*$, and $t_c$ is too weak. To observe cleanly the two
different regimes the time scales must be well separated: $t_\times
\ll t^* \ll t_c$ and this requires very large values of $N$.  Two
additional elements make the observation of the expected scaling of
type I, $\theta=1/(\gamma-2)$, even more difficult.  First, the time
for the onset of scaling (see Appendix~\ref{sec:macr-time-scale})
\begin{equation}
t_\times = \frac{\rho_0^{2-\gamma}
  (\gamma-1)^{\gamma-2}}{(\gamma-2)^{\gamma-1}
  \Gamma(3-\gamma)\Gamma(\gamma-1)}, 
\end{equation}
can be quite large an it diverges for $\gamma \to 2$. Second, the very
slow convergence of $\Theta[\rho]$ to its asymptotic shape
(Fig.~\ref{drift}). Despite these difficulties,
Fig.~\ref{effectiveexp} suggests that by increasing the size of the
system we should eventually be able to recover the theoretical exponent
$\theta=1/(\gamma-2)$

Concerning the exponent corresponding to the type II drift,
$\theta=1$, its evaluation from Fig.~\ref{effectiveexp} is more
difficult. For instance, in the case $\omega=\gamma-1$ the scaling of
$t^*$ and $t_c$ is the same and, therefore, the exponent $\theta=1$
can barely be observed. In the case $\omega=2$, they scale as $t^*
\sim N^{(\gamma-2)/2}$ and $t_c \sim N^{(\gamma-1)/4}$. Their ratio then
goes as $t^*/t_c \sim N^{(\gamma-3)/4}$ ($1/8$ for $\gamma=2.5$) which
is a very small exponent. The direct consequence is that the
evaluation of the exponent $\theta=1$ from these type of experiments
is too influenced by the effect of the crossover between region I and
II.

\begin{figure}
\begin{center}
\includegraphics[width=\columnwidth]{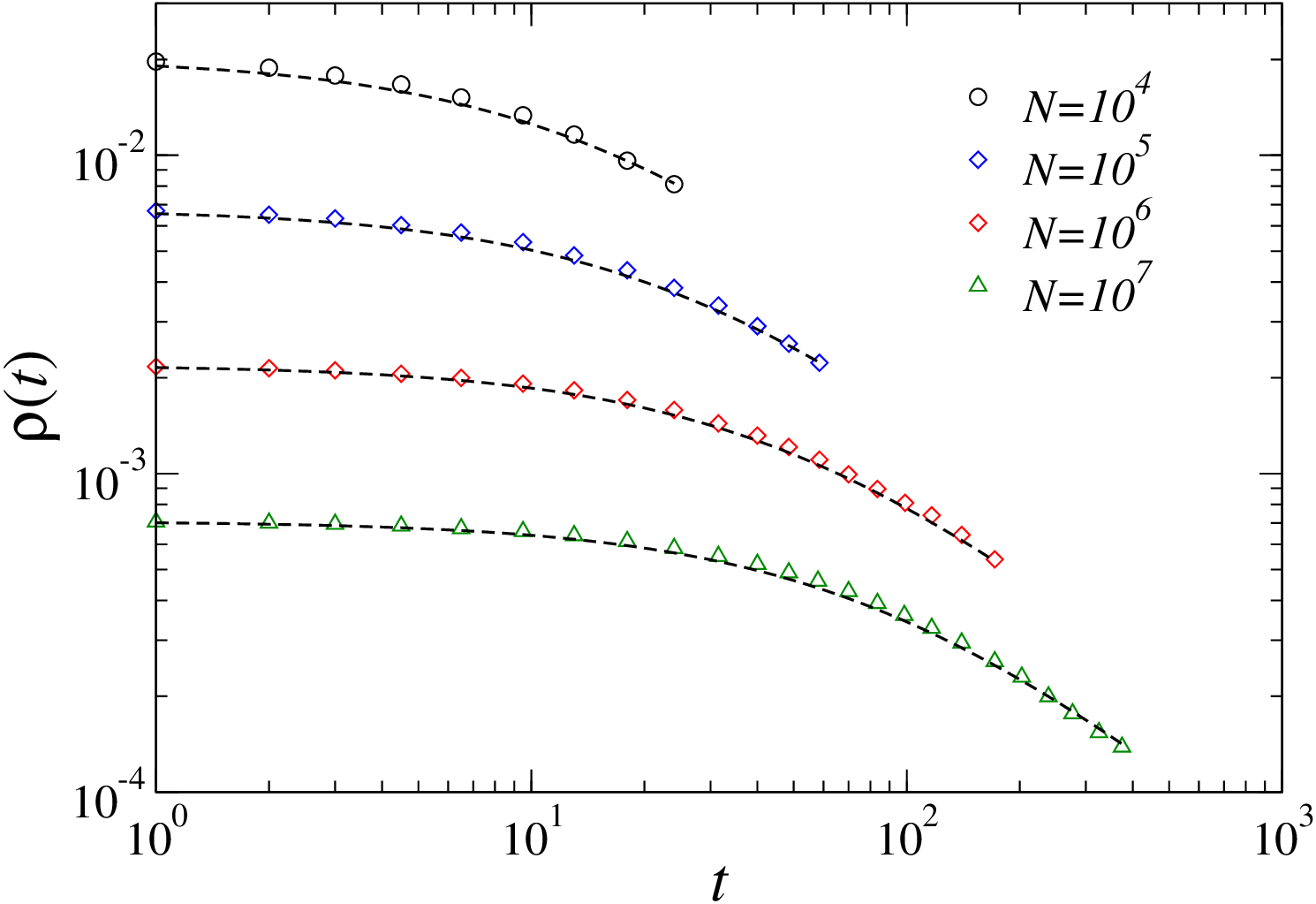}
\end{center}
\caption{Evolution of the concentration $\rho(t)$ in region II for
  different network sizes, $\gamma=2.5$, and $\omega=2$. The initial concentration is
  $\rho_0=\langle k \rangle/2 k_c$ and results are averaged over $500$
  realizations. Dashed lines correspond to the theoretical
  prediction Eq.~(\ref{typeII}).}
\label{rho_type_II}
\end{figure}

To clearly see the predicted behavior corresponding to the type II
drift, we perform numerical simulations with an initial concentration
$\rho_0$ well below the critical level separating regions I and II. In
particular we choose $\rho_0=\langle k \rangle/2 k_c$. With this
initial conditions the dynamics is ruled by the type II drift from the
very beginning, leading to the prediction
\begin{equation} 
\label{typeII}
\rho_{II}(t)=\frac{1}{gt+\rho_0^{-1}}.
\end{equation}
In Fig.~\ref{rho_type_II}, we show simulation results for different
network sizes as compared to the prediction given by
Eq.~(\ref{typeII}). The agreement is very good if we consider that the
dashed lines in Fig.~\ref{rho_type_II} are generated without fitting
any parameter but using the values of $g$ and $\rho_0$ used in the
simulations.

The conclusion is that observing in simulations the exponent
$\theta=1/(\gamma-2)$ predicted by the mean-field theory in the
thermodynamic limit is, although in principle possible as a
pre-asymptotic regime, too difficult from a practical point of view,
since one should reach network sizes that are beyond the
capabilities of current computers.  From Fig.~\ref{effectiveexp} one
can estimate that in order to reach an effective exponent close to 2 a
network larger than $N \approx 10^{11}$ should be considered. On the
other hand, the behavior predicted by type II drift spans for a
shorter time as compared to the type I but is, nevertheless, clearly
visible, as shown in Fig.~\ref{rho_type_II}.

\subsection{Spreading experiments}

Starting from a single randomly chosen active vertex, the activity is
followed until it decays into the absorbing state and the survival
time $t$ is recorded. The survival probability $S(t)$, defined as the
probability that activity lasts longer than $t$, behaves at the
critical point as given by Eq.~(\ref{survival:1}).
 
Fig.~\ref{experiments} shows examples of single realizations of this
experiment (bottom blue curves). In this example, though, we have
selected realizations that survived a time longer than $10^3$, which
roughly corresponds to the value of the cutoff time $t_c$ for this
particular $\gamma$ and $N$. In this way we can see the domain of
$\rho$-space that is visited by the trajectories of the
experiment. Both for $\omega=\gamma-1$ and $\omega=2$,
trajectories never reach the white area, where the type I drift is
dominant. They always remain in the domain governed by the type II drift.

The result of Eq.~(\ref{survival:1}) can be derived from the Langevin
Eq.~(\ref{Langevin:3}), using standard techniques of stochastic
processes theory~\cite{Gardiner:2004}. In
Appendix~\ref{sec:surv-prob-equat}, we show that, in the limit of an
infinite network size, we have
\begin{equation}
  S(t)=\lim_{N\to\infty}S(t,N) =1-e^{-1/t} \approx \frac{1}{t},
\end{equation}
that is, we recover and exponent $\delta=1$ for any degree exponent
$\gamma$.

The value of this exponent implies that the probability density
function of survival times in infinite systems follows a power law of
the form $\psi(t) \sim t^{-2}$ and, therefore, has diverging
fluctuations.  However, in finite size systems, this distribution has
a size-dependent cutoff time $t_c(N)$, and the divergence of the
second moment of survival times $T_2 = \langle t^2 \rangle$ is then
cutoff by $t_c(N)$: $T_2 = 2\int t S(t,N) dt \sim 2 \int^{t_c}
t^{1-\delta} \sim t_c(N)$. A calculation of this second moment
(Appendix~\ref{sec:surv-prob-equat}) leads to the final result
\begin{equation}
t_c(N) \propto \sqrt{\frac{N}{g}}.
\label{t_c}
\end{equation}

This expression has an explicit dependence on the size of the system
but also an implicit one through the size dependence of the factor $g$
that, as we have shown before,
can diverge for $\gamma<3$ with the system size in arbitrary ways.
This, indeed, results in an infinite number of ways to
approach the thermodynamic limit~\cite{Castellano:2008}.
For $\gamma>3$ instead, $g$ is a constant and Eq.~(\ref{t_c}) reproduces
the well-known result of homogeneous MF theory~\cite{Marro}.

\begin{figure}
\begin{center}
\includegraphics[width=\columnwidth]{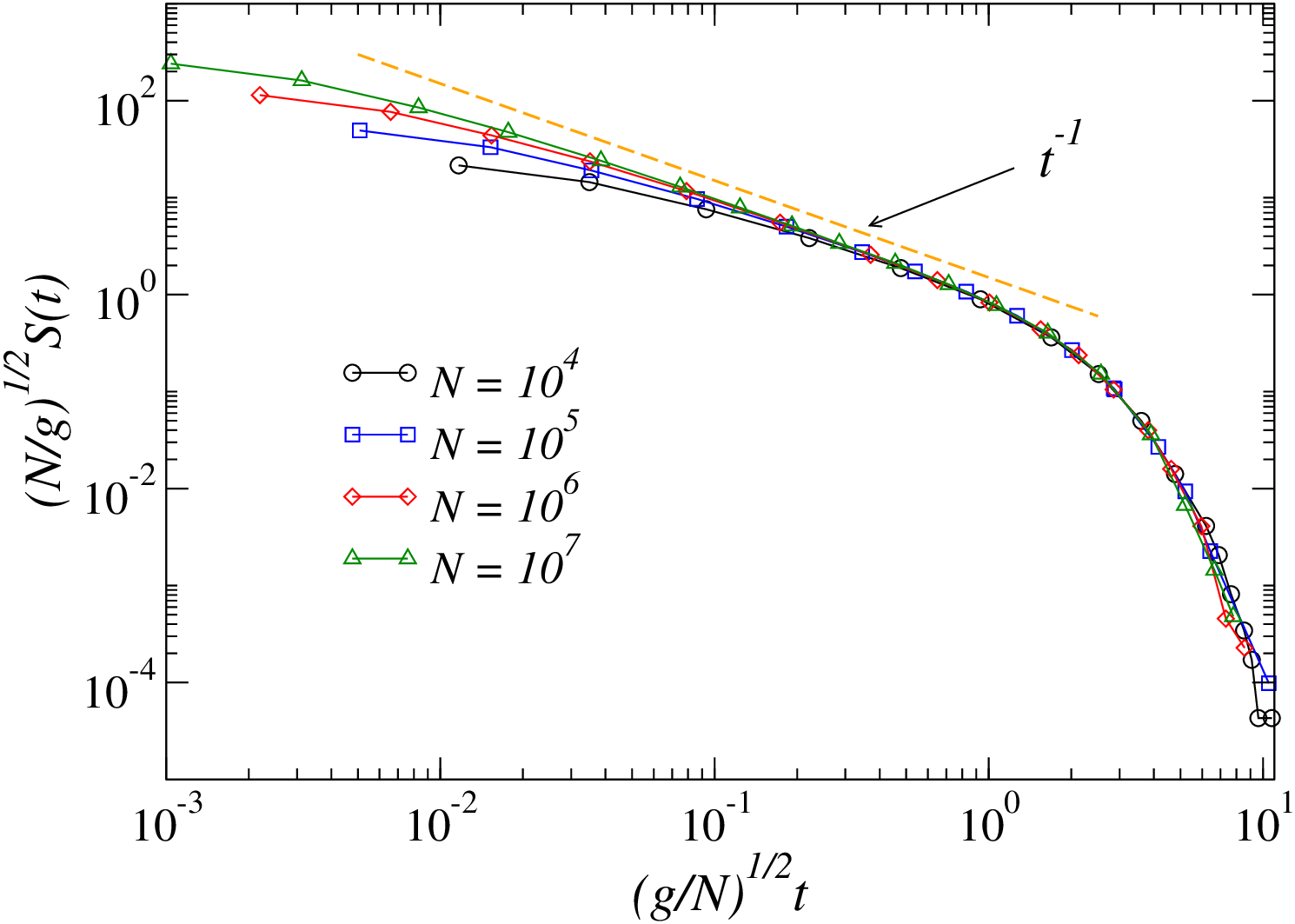}
\end{center}
\caption{Scaling of the survival probability $S(t)$ for $\gamma=2.5$ and
$\omega=\gamma-1$ in a single instance network.}
\label{scaling_omega=gamma-1}
\end{figure}

In Reference~\cite{Castellano:2008} it was shown that
Eq.~(\ref{survival:1}) is obeyed for $\omega=2$ with the scaling
of $t_c(N)$ given by Eq.~(\ref{t_c}), while it is not if
no bound is imposed on the degree distribution.
Fig.~\ref{scaling_omega=gamma-1} shows that the scaling~(\ref{survival:1})
holds also for $\omega=\gamma-1$ with a hard bound.
The violation of the scaling occurring when $\omega<\gamma-1$
has then to do not with the average
value of the maximum degree $k_c\sim N^{1/(\gamma-1)}$ but with the presence
of outliers with exceptionally high values of $k$.

\subsection{Surviving runs}

In this type of experiment, starting from a given initial
concentration, only those trajectories that have survived for a fixed
observation time $T>t_{c}$ are kept and used to compute an average
concentration of active vertices at criticality $\rho_{s}$. From a
numerical point of view, analogous information can be obtained by
means of a surviving average~\cite{Marro}, made over the surviving
representatives of a large number of independent runs.

The motivation for this type of experiment can be traced back to the FSS
theory. According to this phenomenological theory, the concentration
of active vertices in surviving runs satisfies the following scaling
relation~\cite{Marro}
\begin{equation}
  \label{fssother}
\rho_{s}(\Delta, N)=N^{-\beta/\bar{\nu}}f(\Delta N^{1/\bar{\nu}}).
\end{equation}
For SF networks with $2<\gamma<3$, a phenomenological approach in
Ref.~\cite{Hong:2007} predicted $\beta=1/(\gamma-2)$ and
$\bar{\nu}=(\gamma-1)/(\gamma-2)$ (see
Sec.~\ref{sec:finite-size-scaling}). The values of these exponents are
recovered if one considers the type I drift alone. In this section, we
show that, in fact, FSS only hold for heterogeneous networks in the
case of $\omega=\gamma-1$. Even in this case, because of the slow
convergence of the type I drift (e.g. Fig.~\ref{drift}), the FSS
theory presented in~\cite{Hong:2007} can only be observed for
extremely large systems. This goes against the original idea of FSS,
which is used to recover the critical exponents without the need to
reach very large systems.

A critical issue in any FSS theory is the computation of the exponent
$\beta/\bar{\nu}$. According to Eq.~(\ref{fssother}), at the critical point
we expect that the concentration of active vertices in surviving runs
satisfies
\begin{equation}
\rho_{s}(0,N)\sim N^{-\beta/\bar{\nu}}.
\end{equation}
In surviving runs experiments, one selects only those trajectories
that have survived for an arbitrary amount of time. Therefore, the
probability density function that there are $n$ active vertices at
time $t$ restricted only to surviving runs is
\begin{equation}
\label{ps}
p_{s}(n,t|n_{0})=\frac{p(n,t|n_{0})}{S(t|n_{0})},
\end{equation}
where $p(n,t|n_{0})$ is the same probability but measured for all
trajectories, that is, including those that are absorbed at the
boundary. Notice that with this definition $\int dn
p_{s}(n,t|n_{0})=1$. We are interested in the long time limit of this
{\it p.d.f}, $p_{s}(n)=\lim_{t\gg1} p_{s}(n,t|n_{0})$. In this limit,
the concentration of active vertices at criticality for surviving
runs is just
\begin{equation}
\rho_{s}(0,N)=\frac{1}{N} \int n p_{s}(n)dn.
\end{equation}

The probability density function $p(n,t|n_{0})$ satisfies a
Fokker-Planck equation, whose solution allows us to compute the surviving
density (see Appendix~\ref{sec:surv-aver-dens})
\begin{widetext}
\begin{equation}
  \rho_{s}(0,N) \propto \frac{1}{N} \left\{ \sqrt{\frac{\pi N}{2g}}
    \mbox{erf}\left(\sqrt{\frac{\langle k \rangle^2 g
          N}{2k_{c}^2}}\right) +
    \frac{1}{C(\gamma)} \left( \frac{C(\gamma) \langle k \rangle
        N}{\gamma-1}\right)^{\frac{\gamma-2}{\gamma-1}} \Gamma
    \left(\frac{1}{\gamma-1},\frac{C(\gamma) \langle k \rangle
        N}{(\gamma-1)k_{c}^{\gamma-1}} \right) \right\},
  \label{eq:13}
\end{equation}
\end{widetext}
where $ \mbox{erf}(z)$ and $\Gamma(a,z)$ are the error and incomplete
Gamma functions, respectively; the first (second) term in the right
hand side come from type II (type I) drift.  At this point, the result
depends on the particular choice of $\omega$.  Suppose first that
$\omega>\gamma-1$. In this case, both the argument of the error
function and the one of the incomplete Gamma function diverge as $N
\rightarrow \infty$. As a consequence, the contribution of the type I
potential is exponentially small and only the first integral
contributes in the thermodynamic limit, yielding the result
\begin{equation}
\label{rho_s_final}
\rho_{s}(0,N) \propto \frac{1}{\sqrt{g N}}.
\end{equation}
In the case of $\omega=\gamma-1$, the arguments of both the error
function and the incomplete gamma function are constants in the large
size limit. In this case the contribution of both terms is of the same
order in $N$, $\rho_{s}(0,N) \propto N^{-1/(\gamma-1)}$. Nevertheless, since the effective potential $\phi(n,N)$
is a monotonously increasing function of $n$, the contribution in
Eq.~(\ref{eq:13}) of the type I potential is always smaller than that
of the type II.  The physical picture is that trajectories stay most of
the time in region I except for short excursions to region II that
give a small contribution that, nevertheless, is of the same order in
$N$. Therefore, we can conclude that the behavior given by
Eq.~(\ref{rho_s_final}) holds in the whole domain $\omega \in
[\gamma-1,2]$. In terms of $\omega$, we can finally write that
\begin{equation}
\frac{\beta}{\bar{\nu}}=\frac{1}{2} + \frac{3-\gamma}{2\omega}.
\end{equation}
This result implies that the conclusions drawn in
Refs.~\cite{hong07:_comment,Hong:2007} are essentially incorrect,
since the scaling of $\rho_{s}(0,N)$ depends explicitly on the degree
cutoff. The exponent ratio $\beta/\bar{\nu}$ obtained
in~\cite{hong07:_comment,Hong:2007} is
recovered only in the particular case $\omega=\gamma-1$.

\section{The meaning of finite-size scaling}
\label{sec:meaning-finite-size}

With all these results at hand, we can now discuss which is the role,
if any, of FSS theory in the context of absorbing phase transitions in
SF networks. The aim of the FSS ansatz is to connect the
behavior of the system in the active phase---which is independent of
the size of the system---for $\Delta \gg N^{1/\bar{\nu}}$ and the
absorbing one---where there is an explicit size dependence---for
$\Delta \ll N^{1/\bar{\nu}}$. However, the ability to do so relies
upon the ``natural'' assumption that the laws ruling the system do not
change when one performs such transition. In the case of the CP
dynamics in SF networks, we have shown that in the absorbing phase the
system is mainly ruled by type II drift. However, when $\Delta$ is
increased, the concentration of active vertices also increases and
eventually the system starts feeling the type I drift. We are then in
a situation where there is a change of the underlying laws between the
active and absorbing phases. Consequently, FSS theory does not work in
this case. Nevertheless, there are some subtle details depending on
the type of cutoff that we discuss next.

\begin{figure}
\begin{center}
\includegraphics[width=\columnwidth]{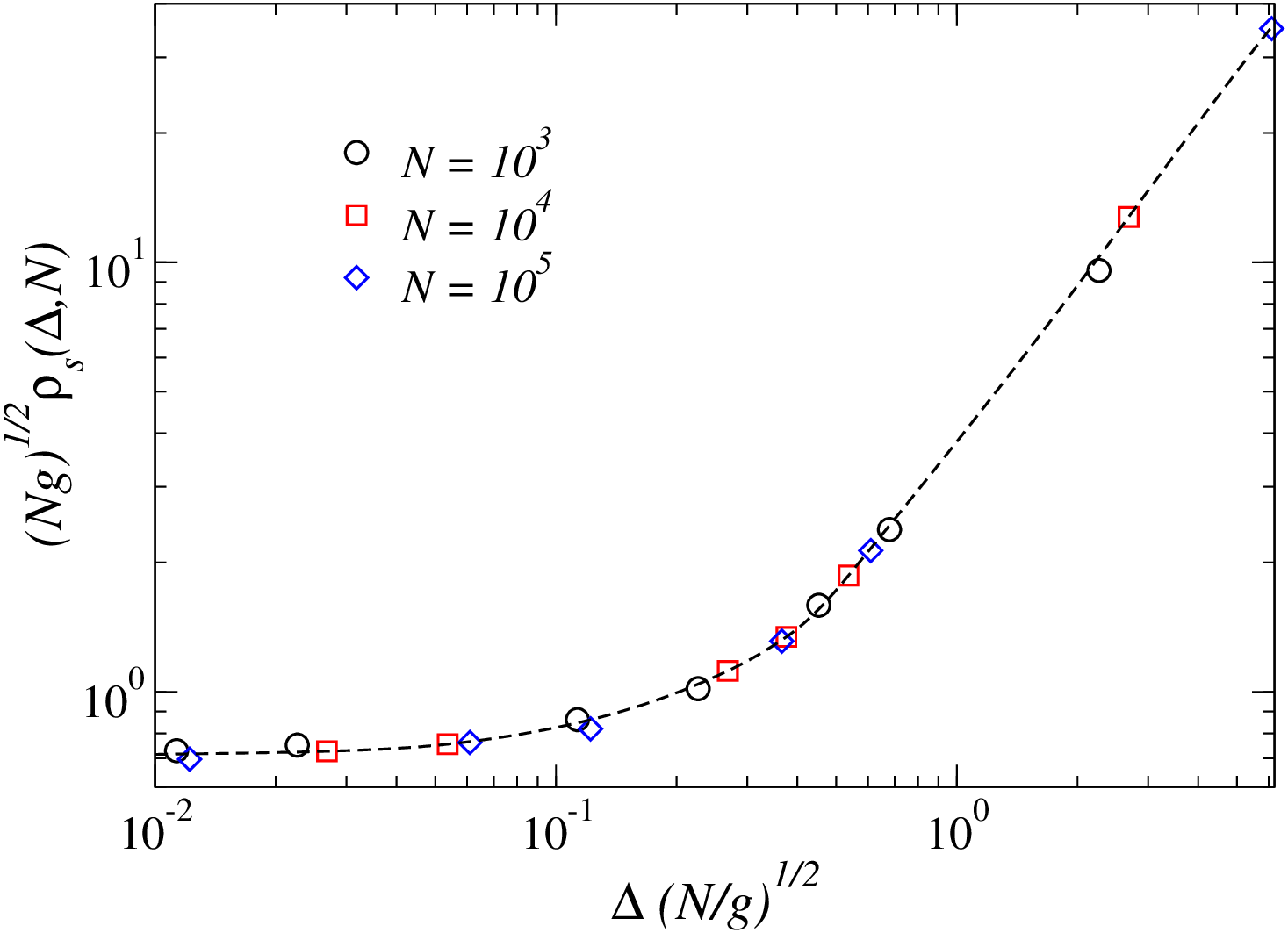}
\end{center}
\caption{Scaling of the density for surviving runs for $\gamma=2.2$ and
$\omega=2$. Each point is the result $10^2$ realizations of the stochastic
process on each of the $10^2$ network realizations. The dashed line is an interpolation of the data as a guide to the eye.}
\label{FSS}
\end{figure}

In the case of $\omega \geq \gamma-1$, the Langevin equation describing
the dynamics for the concentration $\rho$ is
\begin{equation}
\label{langevin_typeII}
  \frac{d \rho(t)}{dt}=\rho(t)\left(\Delta-\lambda^2 g \rho(t)
  \right)+\sqrt{\frac{2\lambda \rho(t)}{N}} \xi(t) 
\end{equation}
that holds if $\Delta \ll \lambda g \langle k \rangle/k_{c}$. If we
perform the change of variables $N_{ef}=N/g$ and $\rho_{ef}=n/N_{ef}$,
the previous equation becomes
\begin{equation}
  \frac{d \rho_{ef}(t)}{dt}=\rho_{ef}(t)\left(\Delta-\lambda^2
    \rho_{ef}(t) \right)+\sqrt{\frac{2\lambda \rho_{ef}(t)}{N_{ef}}} \xi(t).
\end{equation}
Notice that this equation describes the CP dynamics in a homogeneous
network of effective size $N_{ef}$. Therefore,
$\rho_{ef}(\Delta,N_{ef})$ must satisfy a FFS with exponents $\beta=1$
and $\bar{\nu}=2$, that is
\begin{equation}
  \rho_{ef}(\Delta,N_{ef})=\frac{1}{\sqrt{N_{ef}}}f(\Delta
  \sqrt{N_{ef}}). 
\end{equation}
Undoing the change of variables we conclude that $\rho_{s}(\Delta,N)$
satisfy the anomalous FSS
\begin{equation}
  \rho_{s}(\Delta,N)=\frac{1}{\sqrt{gN}} f\left(\Delta
    \sqrt{\frac{N}{g}}\right) \mbox{\hspace{0.2cm} for \hspace{0.2cm}}
  \Delta \ll \frac{\lambda g \langle k \rangle}{k_{c}}.
\label{eqFSS}
\end{equation}
This FSS is anomalous in the sense that when $\Delta>\sqrt{g/N}$ then
$\rho_{s}(\Delta,N) \sim \Delta/g$ which depends on the size of the
system through the factor $g$. Notice also that for $\gamma<3$ the factor $g/k_{c}$ can be
reasonably large even for large system sizes and, consequently, this
anomalous scaling can be observed in a wide range of values of
$\Delta$.
Fig.~\ref{FSS} confirms the validity of Eq.~(\ref{eqFSS}).

\section{The effect of outliers}
\label{sec:effect-outliers}

In the previous sections, we have assumed that the maximum allowed
degree of the network $M$ scales with the system size $N$ as
$N^{1/\omega}$, with $\omega \ge \gamma-1$, so that the average cutoff
in the degree distribution $k_c$ is proportional to $M$ and degrees
much larger than $k_c$ are simply forbidden.

When $M$ scales faster than $N^{1/(\gamma-1)}$ instead, the cutoff
degree $k_c$ is not a hard but a ``soft'' statistical value: the
maximum degree $\kmax$ in a single realization of the network is the
result of a random process that yields $k_c$ on average but has
diverging fluctuations: it is
still possible to find outlier vertices having degrees much larger
that $k_c$. To investigate which is the role of outliers in the CP
dynamics, we introduce a minimal toy network model with a hard cutoff
$k_c \le N^{1/(\gamma-1)}$, just as in the previous sections, and then we add
a single vertex of degree $k_{out}=\alpha N$, with $\alpha \in [0,1]$.

From Eq.~(\ref{adiabatic}) we see that the concentration in
the outlier vertex is $\rho_{k_{out}} \approx 1$ provided that
$\rho \gg \avk /(\lambda \alpha N)$,
where the average degree must be computed including the contribution
of the outlier.
This condition is satisfied both in density decay and surviving runs
experiments. In the case of spreading experiments, the condition is
satisfied only partially since during the beginning of the experiment
the density is always of the order $\rho \sim N^{-1}$.
However, in the first two types of experiments, the effect of the
outlier vertex is that the rest of the vertices ``see'' the outlier
always active. Since the outlier holds a macroscopic portion of the
edges of the system, all attempts to make it active
occurring along one of its edges are unsuccessful. The net effect
is that the system is shifted away from its critical point $\lambda=1$
and is effectively in a sub-critical state.
To quantify this effect and to calculate the position of the new critical
point, we separate in Eq.(\ref{eq:7}) the outlier's contribution
from that of the rest of the vertices. This results in
\begin{equation}
  \label{theta_outlier}
  \Theta[\rho(t)]= \sum_{k\neq k_{out}}^{k_c}
  \frac{kP(k)}{\langle k \rangle}\frac{\lambda k \rho(t)/\langle k
    \rangle} 
  {1+\lambda k \rho(t)/\langle k \rangle}+\frac{k_{out}}{N \langle k
    \rangle}, 
\end{equation}
that is, the outlier has a constant contribution to the drift term
whereas the contribution of the rest of the vertices goes to zero when
$\rho$ approaches zero. Combining this result with
Eq.~(\ref{Langevin:3}) we obtain the new critical point
\begin{equation}
  \lambda_c^0=\frac{1}{1-k_{out}/N \avk}.
\label{lambda_eff}
\end{equation}

\begin{figure}
\begin{center}
\includegraphics[width=\columnwidth]{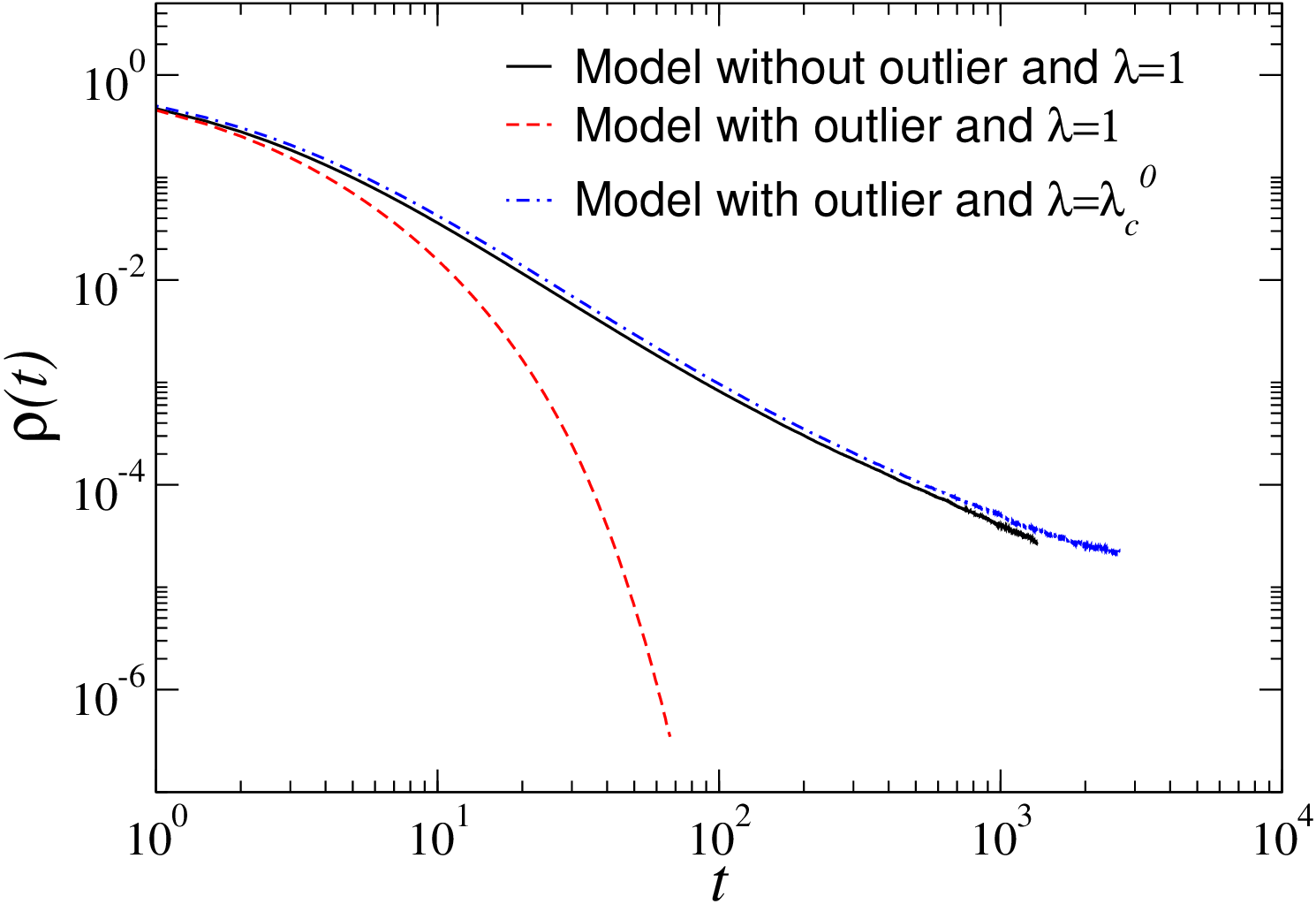}
\end{center}
\caption{Density decay starting from a fully active network for the
  model with and without outlier. Network size is $N=10^8$,
  $\gamma=2.5$, and $M=N^{1/2}$. Results are
  averaged over $100$ realizations in a
  single instance network.}
\label{outlier}
\end{figure}

To check this result, we perform numerical simulations starting from a
fully active network in three different scenarios (see Fig.~\ref{outlier}).
In the first one, we generate a network with
$M=N^{1/2}$ and set the dynamics to its critical value
$\lambda=1$. Here, $\lambda=1$ is the true critical point and, as
expected, we find a double power law decay towards the absorbing
state, as explained in section \ref{sec:CPSF}.  In the second
scenario, we introduce in the previous network a single vertex of
degree $k_{out}=N$ and, again, set $\lambda=1$. In this case, we
observe a clear exponential decay, typical of a sub-critical
regime. Finally, in the third experiment, we keep the network with the
outlier but we increase the control parameter according to
Eq. (\ref{lambda_eff}). After this correction to the critical point,
we observe again a clear double power law decay towards the absorbing
state, indicating that, indeed, Eq. (\ref{lambda_eff}) predicts the
correct critical point $\lambda_c^0$.

\begin{figure}
\begin{center}
\includegraphics[width=\columnwidth]{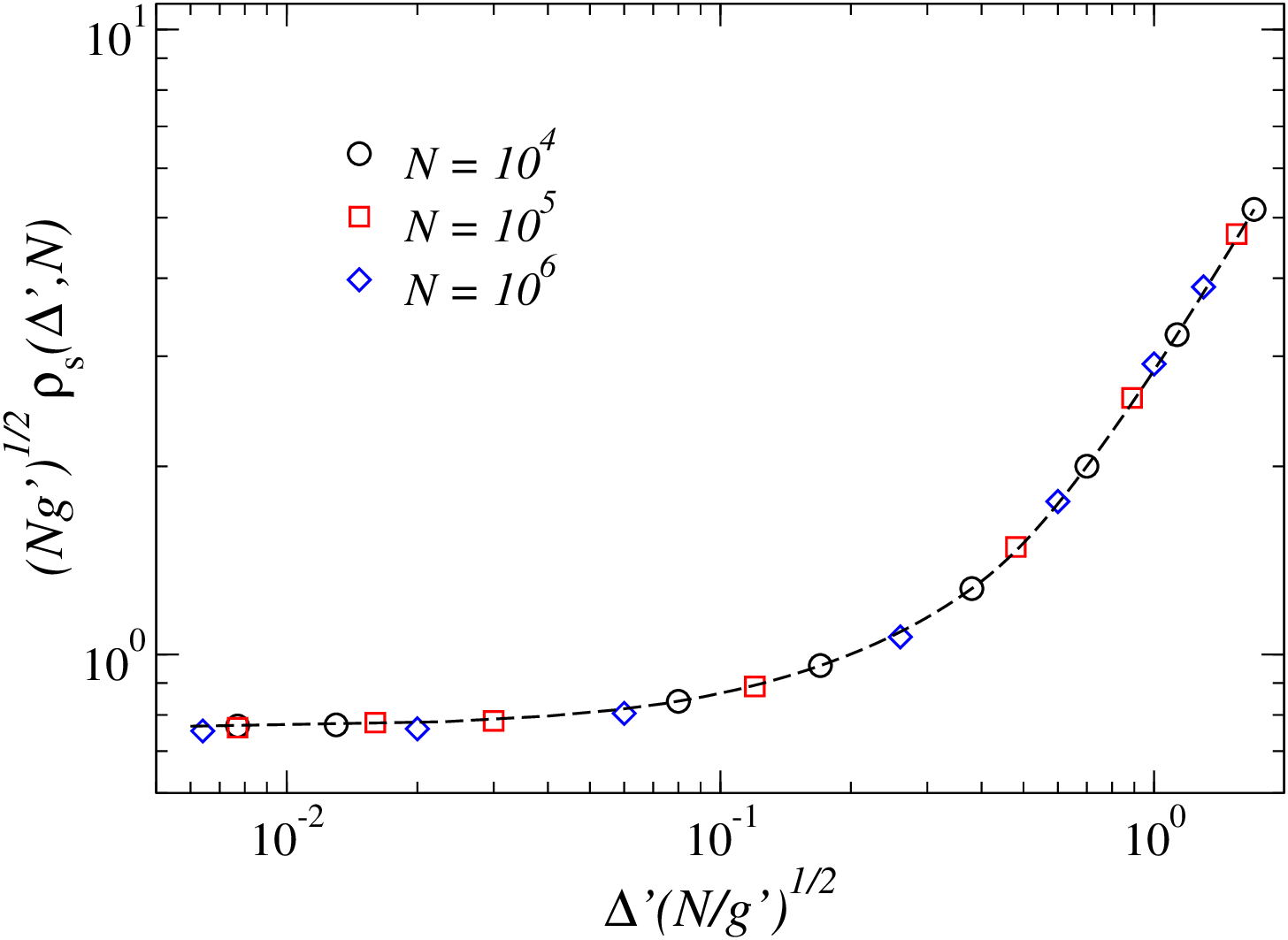}
\end{center}
\caption{Scaling of the density for surviving runs for $\gamma=2.5$
  and $\omega=2$, with at outlier of degree $k_{out}=N$. Each point is
  the result $10^4$ realizations of the stochastic process on a single
  network realization. The dashed line is an interpolation of the data as a guide to the eye.}
\label{outlier3}
\end{figure}

Along the same lines it is possible to understand also surviving runs
in the same network with an outlier.  The equation of motion for the
density $\rho$ at the new critical point $\lambda=\lambda_c^0$ is
\begin{equation}
  \frac{d \rho(t)}{dt}=-(\lambda_c^0)^2 g' \rho^2(t)
   +\sqrt{\frac{2 \rho(t)}{N}} \xi(t),
   \label{eq:11}
 \end{equation}
where 
\begin{equation}
  g' = \sum_{k \ne k_{out}} k^2 P(k)/\avk^2.
  \label{eq:12}
\end{equation}
After redefining time in Eq.~(\ref{eq:11}) as $t'=(\lambda_c^0)^2 t$,
we recover the same type of Langevin equation as for the
case $\omega \geq \gamma-1$, Eq.~(\ref{langevin_typeII}), but with an
effective parameter $g'$
given by Eq.~(\ref{eq:12}). From here, we readily obtain the FSS form for
the average density in surviving experiments, namely
\begin{equation}
  \rho_{s}(\Delta',N)=\frac{1}{\sqrt{g' N}} f\left(\Delta'
    \sqrt{\frac{N}{g'}}\right),
\label{eqFSSOut}
\end{equation}
where $\Delta' = \lambda-\lambda_c^0$.
In Fig.~\ref{outlier3} the validity of the scaling form given
by Eq.~(\ref{eqFSSOut}) is demonstrated.

\section{Conclusions}
\label{sec:conclusions}

In this paper, we have presented a detailed analysis of the
dynamics of the contact process on annealed scale-free networks. Using
stochastic differential equations for this dynamics, we have clarified
the behavior of the model close to its critical point and, in
particular, its finite size scaling. Our results indicate that
heterogeneous mean-field theory---strictly valid for infinite
networks---is practically unobservable for the range of sizes that
modern computers can reach. The dynamics is instead dominated by
strong finite size effects that give rise to nontrivial anomalous
effects.  Among them, it is worth to notice that the scaling of
several relevant quantities (like the order parameter close to the
critical point or the surviving times of the dynamics, etc) depends
not only on the system size $N$, as in regular lattices, but also on
the upper cutoff $M$ of the scale-free degree distribution which, in
general, diverges with the system size as $M \sim N^{1/\omega}$. The
exponent $\omega$ is not fixed by the degree distribution alone and,
in general, can take different values for different network models or
even an arbitrary value that we can freely choose in annealed
networks. This implies that the critical exponents of the dynamics are
not universal but depend on the arbitrary value of $\omega$.

Our results allow us to understand the origin of the discrepancy
between the phenomenological finite size scaling theory proposed in
Ref.~\cite{Hong:2007} and the numerical results found
in~\cite{castellano07:_reply,Castellano:2008}.  Indeed, the Langevin
equation proposed in Ref.~\cite{Hong:2007}, Eq.~(\ref{KoreanLE}),
misses the crucial point that the coefficient $b$ is not a constant
but depends on the system size, and grows with it. As a consequence,
when the system is at its critical point and the concentration becomes
small enough, the term $b \rho^2$ in Eq.~(\ref{KoreanLE}) becomes more
important than the term $d \rho^{\gamma-1}$, something that would not
be possible if $b$ was a constant and $\gamma<3$. This change in the
dominating term in Eq.~(\ref{KoreanLE}) at low concentrations
invalidates thus the FSS proposed in~\cite{Hong:2007}.

When $\omega<\gamma-1$ an additional interesting complication arises:
the effective average cutoff of the degree distribution $k_c$ becomes
$N^{1/(\gamma-1)}$, smaller than $M$, while its fluctuations diverge
as $N$ grows. This means that, depending on the specific realization
of the degree sequence, some outliers (i.e. nodes with a connectivity
much larger than the effective average cutoff $k_c$) may appear.
We have shown that a single
outlier connected to a macroscopic portion of the system has the
effect of introducing an apparent shift on the critical point
position.  The investigation of the role of outliers (and more in
general the role of diverging fluctuations in the effective upper
cutoff) in the contact process and other models is a very interesting
avenue for further investigations. Notice that this case is the
relevant one for simulations performed without fixing an explicit
upper cutoff of the degree distribution, a very common habit.

Last but not least, we would like to stress that the theory and the
simulations presented here give a complete understanding of the
complex behavior of the contact process on {\em annealed} scale-free
networks.  Whether or not the same picture also holds for {\em
  quenched} topologies remains an open question calling for further
work.

\begin{acknowledgments}
R.~P.-S.  and M.~B. acknowledge financial support from the Spanish MEC (FEDER) under projects FIS2007-66485-C02-01 and FIS2007-66485-C02-02. R.~P.-S. also acknowledges the hospitality of the Institute for Scientific Interchange Foundation, Turin (Italy), where part of this work was developed.
\end{acknowledgments}

\vspace*{1cm}
\begin{appendix}

\section{Calculation of the coarse-grained Langevin equation}
\label{sec:appendix-calc-coarse-grain}

The derivation of the Langevin equation that describes a general
stochastic process $X(t)$ involves the evaluation of its infinitesimal
moments. The first and second (variance) such moments inform us about
the expected change in the process after an increment of time $dt$ and
the variance of this expected change. More precisely, if we define the
variable $\Delta X(t) \equiv X(t+dt)-X(t)$~\cite{Gardiner:2004}, then
the first infinitesimal moment is defined as
\begin{equation}
\Psi[x]=\lim_{dt\rightarrow 0} \frac{\langle \Delta X(t) |X(t)=x\rangle}{dt}.
\end{equation}
Analogously, the infinitesimal variance is defined as
\begin{equation}
D[x]=\lim_{dt\rightarrow 0} \frac{\langle [\Delta X(t)]^2 |X(t)=x\rangle}{dt}.
\end{equation} 
Functions $\Psi[x]$ and $D[x]$ are called the drift and the diffusion
term, respectively. The Langevin stochastic differential equation can
then be written as
\begin{equation}
\frac{dX(t)}{dt}=\Psi[X(t)]+\sqrt{D[X(t)]}\xi(t),
\end{equation}
where $\xi(t)$ is a Gaussian white noise.

In our case, we are interested in writing a Langevin equation for the
coarse-grained quantity $n_k(t)=\sum_{i\in k}\sigma_i(t)$. This is
convenient for two main reasons: First, since it is the sum of almost
(or totally) independent random variables, we expect the central limit
theorem to hold. This guarantees that the corresponding noise in the
Langevin equation is Gaussian and white. Second, in the thermodynamic
limit, $n_k(t)/N$ can be safely assumed to be a continuous variable
and, therefore, it is justified the use of (stochastic) differential
equations.

To compute the infinitesimal first moment we write
\begin{equation}
 \langle n_k(t+dt) | {\bf \Sigma}(t) \rangle=\sum_{i \in
   k}\sigma_i(t)\langle \zeta_i(dt)\rangle+[1-\sigma_i(t)]\langle
 \eta_i(dt)\rangle
\end{equation}
where we have made use of Eq.~(\ref{micro}). Finally, using the
probability distributions Eqs.~(\ref{xi}) and~(\ref{eta}), we are lead to
\begin{widetext}
\begin{equation}
  \label{first_moment}
  \langle n_k(t+dt) | {\bf \Sigma}(t) \rangle = n_k(t) +dt \left[
    -n_k(t) +\lambda \sum_{k'}\frac{1}{k'} \sum_{i\in ks; j \in k'}
    a_{ij} \left[1-\sigma_i(t)\right]\sigma_j(t) 
  \right].
\end{equation}
Analogously, we can write an expression for the infinitesimal variance as
\begin{equation}
  \label{second_moment}
  \langle n_k^2(t+dt) | {\bf \Sigma}(t) \rangle-\langle n_k(t+dt) | {\bf
    \Sigma}(t) \rangle^2   =dt \left[  n_k(t)+\lambda \sum_{k'}\frac{1}{k'}
    \sum_{i\in k ; j \in k'} a_{ij} \left[1-\sigma_i(t)\right]\sigma_j(t) 
  \right].
\end{equation}
\end{widetext}
To derive the previous equation, we have taken into account that,
since $\sigma_i(t)$ are binary variables taking only values $0$ or $1$,
$\sigma_i^2(t)=\sigma_i(t)$ and that $\sigma_i(t)[1-\sigma_i(t)]=0; \forall t$.
Terms of order $dt^2$ have also been neglected. Under the annealed
approximation, we replace in Eqs.~(\ref{first_moment}) and
(\ref{second_moment}) the adjacency matrix $a_{ij}$ by its average
value, Eq.~(\ref{eq:3}), which allows us to carry out the sums in
Eqs.~(\ref{first_moment}) and (\ref{second_moment}) and, finally, to
obtain the Langevin equation Eq.~(\ref{Langevin:1}).

\section{Calculation of $\Theta[\rho]$  in uncorrelated SF networks}
\label{sec:calculation-thetarho}

Let us consider the definition of $\Theta[\rho]$ in Eq.~(\ref{eq:7}),
namely
\begin{equation}
  \Theta[\rho]= \sum_k \frac{kP(k)}{\langle k
    \rangle}\frac{\lambda k \rho/\langle k \rangle}{1+\lambda k
    \rho/\langle k \rangle}. 
  \label{eq:8}
\end{equation}
The evaluation of this quantity in finite networks depends on the
value of the particle density $\rho$. In particular, if $\rho \ll
\avk/ \lambda k_c$, where $k_c$ is the network cutoff, then the
denominator in Eq.~(\ref{eq:8}) can be approximated by unity, and we
have\begin{equation}
  \Theta[\rho] \simeq  \sum_k \frac{kP(k)}{\avk}\frac{\lambda k
    \rho}{\avk} = \frac{\fluck{k}}{\avk} \frac{\lambda 
    \rho}{\avk} = g \lambda \rho,
\end{equation} 
where $g=\fluck{k}/\avk^2$.  On the other hand, outside this region we
must keep the full denominator in Eq.~(\ref{eq:8}). To estimate
$\Theta$ in this case, we perform a continuous degree approximation, that is,
 \[
  \Theta[\rho]= \int_m^{k_c}  \frac{kP(k)}{\langle k
    \rangle}\frac{\lambda k \rho/\langle k \rangle}{1+\lambda k
    \rho/\langle k \rangle}
\]
\begin{equation}
\footnotesize
  = F\left[1, \gamma-2, \gamma-1, -\frac{\avk}{\lambda \rho m} \right]-
  F\left[1, \gamma-2, \gamma-1, -\frac{\avk}{\lambda \rho k_c}
  \right] \left(\frac{m}{k_c}\right)^{\gamma-2},
 \label{theta_approx} 
\end{equation}
where $F[a,b,c,z]$ is the Gauss hypergeometric function. Using the
asymptotic expansions of the hypergeometric function for small and
large arguments~\cite{abramovitz}, we can estimate the value of
$\Theta$ in the domain $\avk/ \lambda k_c \ll \rho \ll 1$. Within such
domain, the second term in Eq.~(\ref{theta_approx}) becomes an
asymptotically small constant as compared to the first term, that
yields\begin{equation} \Theta[\rho] \simeq \Gamma(\gamma-1)
\Gamma(3-\gamma) \left( \frac{\lambda \rho m}{\avk}
\right)^{\gamma-2}.
\end{equation}

\section{Initial time scale}
\label{sec:macr-time-scale}

To compute the initial time scale $t_\times$ needed to reach
region I starting from an arbitrary initial condition $\rho_0$ at
criticality, we consider Eq.~(\ref{Langevin:3}) with the drift term
given by Eq.~(\ref{driftII}) and $\Delta=0$, namely
\begin{equation}
  \frac{d \rho(t)}{dt}= -\frac{C(\gamma)}{\langle
      k  \rangle^{\gamma-2}} \rho(t)^{\gamma-1}.
\end{equation}
The solution of this equation is
\begin{equation}
   \rho(t)=  \left[\rho_0^{2-\gamma} + \frac{(\gamma-2)
     C(\gamma)}{\langle       k  \rangle^{\gamma-2}}\;
   t\right]^{-1/(\gamma-2)}. 
\end{equation}
The asymptotic state $\rho(t) \sim t^{-1/(\gamma-2)}$, independent of
the initial condition, is reached for times $t$, such that 
\begin{equation}
  \frac{(\gamma-2)C(\gamma)}{\langle k \rangle^{\gamma-2}}\;  t \gg
  \rho_0^{2-\gamma},
\end{equation}
that is, for $t> t_\times$, with
\begin{equation}
  t_\times = \frac{ \rho_0^{2-\gamma} \langle k
    \rangle^{\gamma-2}}{(\gamma-2)C(\gamma)}=\frac{\rho_0^{2-\gamma}
  (\gamma-1)^{\gamma-2}}{(\gamma-2)^{\gamma-1}
  \Gamma(3-\gamma)\Gamma(\gamma-1)},
\end{equation}
where we have used the definition of $C(\gamma)$ in Eq.~(\ref{eq:9})
and $\avk = (\gamma-1) m / (\gamma-2)$.

\section{Survival probability equation}
\label{sec:surv-prob-equat}

Using standard techniques of stochastic processes theory, we can
obtain the partial differential equation satisfied by the survival
probability of the CP dynamics at criticality, starting from an initial
concentration $\rho_0$, $S(t|\rho_0)$, namely~\cite{Gardiner:2004}
\begin{equation}
  \label{partial_survival:1}
  \frac{\partial S(t|\rho_0)}{\partial t}=-\rho_0 \Theta[\rho_0]
  \frac{\partial S(t|\rho_0)}{\partial \rho_0}+\frac{\rho_0}{N}
  \frac{\partial^2 S(t|\rho_0)}{\partial \rho_0^2}.
\end{equation}
This equation is the result of integrating the backwards Fokker-Planck
equation in the domain $\rho \in [0,1]$ and it should be solved with
the initial condition $S(t=0|\rho_0)=1$ and boundary conditions
\begin{equation}
  S(t|\rho_0=0)=0 \mbox{\hspace{0.2cm}
    and\hspace{0.2cm}}\left. \frac{\partial S(t|\rho_0)}{\partial
      \rho_0}\right|_{\rho_0=1}=0 
\end{equation}
that correspond to an absorbing boundary at $\rho=0$ and a reflecting
one at $\rho=1$.  The survival probability Eq.~(\ref{survival:1}) can
then be evaluated as
\begin{equation}
  S(t)=S(t|\rho_0=1/N).
\end{equation}

We first start by evaluating the exponent $\delta$. To this end, it is
only necessary to solve the problem in the thermodynamic limit
$N\rightarrow \infty$. However, the above formulation is not the most
appropriate for this purpose, since the solution must be evaluated at
$\rho_0=N^{-1}$, that is, a value that depends on the size of the
system. Therefore, we perform the change of variables
\begin{equation}
n_0=N \rho_0
\end{equation}
where $n_0$ is the initial number of active vertices, which is
eventually set to $n_0=1$ and, therefore, is independent of the system
size. Using this new variable, Eq.~(\ref{partial_survival:1}) becomes
\begin{equation}
  \label{partial_survival:2}
  \frac{\partial S(t|n_0)}{\partial t}=-n_0
  \Theta\left[\frac{n_0}{N}\right] \frac{\partial S(t|n_0)}{\partial
    n_0}+n_0 \frac{\partial^2 S(t|n_0)}{\partial n_0^2}. 
\end{equation}
Notice that now the limit $N\rightarrow \infty$ can be taken in
Eq.~(\ref{partial_survival:2}). In this limit, the first term in the
right hand side of Eq.~(\ref{partial_survival:2}) vanishes and the
process becomes a purely diffusive one with multiplicative noise. The
solution is
\begin{equation}
\label{survival_prob}
S(t|n_0)=1-e^{-n_0/t} \approx \frac{n_0}{t}.
\end{equation}
Setting finally $n_0=1$, leads to the exponent $\delta=1$ for any
$\gamma$.

To evaluate the cutoff $t_c(N)$, we compute the second moment of the
survival times, $T_2(n_0)$, starting from $n_0$ active sites. However,
to compute $T_2$ we first need to compute the average surviving time
$T_1(n_0)$. It is easy to see that $T_1(n_0)=\int_0^{\infty}
S(t|n_0)$. Using this result in~Eq.(\ref{partial_survival:2}), and
assuming that trajectories never feel the type I drift, yields the
following differential equation for $T_1(n_0)$
\begin{equation}
\frac{d^2 T_1(n_0)}{dn_0^2}-\frac{g}{N}n_0 \frac{dT_1(n_0)}{dn_0}=-\frac{1}{n_0}.
\end{equation}
with boundary conditions $T_1(0)=0$ and $T'_1(N)=0$. The solution of
this problem is
\begin{equation}
T_1(n_0)=\sqrt{\frac{2N}{g}} \int_0^{n_0\sqrt{\frac{g}{2N}}}du e^{u^2}
\int_u^{\sqrt{gN/2}} \frac{dt}{t} e^{-t^2}.
\end{equation}
When $N$ is very large, the upper limit in the first integral becomes
very small. Therefore, we take the limit of the integrand when $u$ is
close to zero, that is,
\begin{equation}
T_1(n_0) \simeq \sqrt{\frac{2N}{g}} \int_0^{n_0\sqrt{\frac{g}{2N}}}du
[1+u^2+\cdots] [-\ln{u}+\gamma+\cdots]
\end{equation}
which finally leads to
\begin{equation}
T_1(n_0) \simeq -n_0 \ln{\left[n_0\sqrt{\frac{g}{2N}}\right]}.
\end{equation}
Similarly, the differential equation for $T_2(n_0)$ can also be
obtained from ~Eq.(\ref{partial_survival:2}) as
\begin{equation} 
T_2(n_0)=2 \int_0^{\infty} t S(t|n_0)dt.
\end{equation}
This results in the following differential equation (involving also $T_1(n_0)$)
\begin{equation}
\frac{d^2 T_2(n_0)}{dn_0^2}-\frac{g}{N}n_0
\frac{dT_2(n_0)}{dn_0}=-\frac{2 T_1(n_0)}{n_0}.
\end{equation}
that satisfies the same boundary conditions as $T_1(n_0)$. The
solution of this equation is
\begin{equation}
  T_2(n_0)=\frac{2N}{g}\int_0^{n_0\sqrt{\frac{g}{2N}}}du e^{u^2}
  \int_u^{\infty} G(t) e^{-t^2} dt 
\end{equation}
where
\begin{equation}
  G(t)=\frac{2}{t} \int_0^t du e^{u^2} \int_u^{\infty} \frac{dq}{q}
  e^{-q^2} 
\end{equation}
In the limit of large $N$, this expression can be approximated as
\begin{equation}
  T_2(n_0)= n_0 \sqrt{\frac{2N}{g}} \int_0^{\infty} e^{-t^2} G(t)dt,
\end{equation}
proving then Eq.~(\ref{t_c}).

\section{Probability density function for surviving runs}
\label{sec:surv-aver-dens}

At the critical point, the probability density $p(n,t|n_{0})$ of the
number of active vertices at time $t$ given that the process had $n_0$
active ones at time $t=0$ is ruled by a Fokker-Planck equation with a
drift term $\Psi(n)=-n \Theta[n/N]$ and a diffusion coefficient
$D(n)=2n$, that is,
\begin{equation}
\label{FP}
\frac{\partial}{\partial n}\left[ n \Theta\left[\frac{n}{N}\right]
  p(n,t|n_0)\right]+ \frac{\partial^2}{\partial n^2}\left[ n
  p(n,t|n_0)\right]=\frac{\partial p(n,t|n_0)}{\partial t}.
\end{equation}
A direct substitution of Eq.~(\ref{ps}) into Eq.~(\ref{FP}) leads to
\begin{widetext}
\begin{equation}
\label{FPII}
\frac{\partial}{\partial n}\left[ n \Theta\left[\frac{n}{N}\right]
  p_s(n,t|n_0)\right]+ \frac{\partial^2}{\partial n^2}\left[ n
  p_s(n,t|n_0)\right]= \frac{\partial p_s(n,t|n_0)}{\partial
  t}+p_s(n,t|n_0)\frac{d \ln{[S(t|n_0)]}}{d t}.
\end{equation}
\end{widetext}
The density $p_s(n,t|n_0)$ has, by construction, a well-defined steady
state, that we denote by
\begin{equation}
p_s(n) \equiv \lim_{t\gg 1} p_s(n,t|n_0),
\end{equation}
which is independent of the initial condition. By taking the limit $t
\gg 1$ in Eq.~(\ref{FPII}), we obtain
\begin{equation}
\frac{\partial}{\partial n}\left[ n \Theta\left[\frac{n}{N}\right]
p_{s}(n)\right]+
\frac{\partial^2}{\partial n^2}\left[ n p_{s}(n)\right]=\kappa p_{s}(n) ,
\end{equation}
where
\begin{equation}
\kappa=\lim_{t\gg 1} \frac{d }{dt} \ln\left[S(t|n_{0})\right]
\end{equation}
Using the result given in Eq.~(\ref{survival_prob}) we conclude that
$\kappa=0$, meaning that $p_{s}(n)$ satisfies the potential solution
of the Fokker-Planck equation~\cite{Gardiner:2004}. We can then write
that
\begin{equation}
\rho_{s}(0,N) \propto \frac{1}{N} \int_{1}^{N} e^{-\phi(n,N)}dn,
\end{equation}
with the effective potential
\begin{equation}
\phi(n,N)=\int \Theta\left[\frac{n}{N}\right] dn.
\end{equation}
As in the case of the function $\Theta$, the potential $\phi(n,N)$
takes a different functional form depending on the value of
$n$. Direct integration of Eq.~(\ref{theta}) gives
\begin{equation}
  \label{potential}
  \phi(n,N)=\left\{
    \begin{array}{lr}
      \frac{\lambda g}{2N}n^2 & n \ll \frac{\langle k \rangle
        N}{\lambda k_c}\\[0.5cm] 
      \frac{C(\gamma)}{\gamma-1} \left( \frac{\lambda}{\langle k
          \rangle N}\right)^{\gamma-2} n^{\gamma-1}&  
      \frac{\langle k \rangle N}{\lambda k_c} \ll n \ll N 
    \end{array}
  \right.
\end{equation}
At the critical point, $\lambda=1$, we can use this result to write 
\begin{equation}
  \label{rho_s}
  \rho_{s}(0,N) \propto \frac{1}{N} \left\{ \int_{1}^{\frac{\langle k
        \rangle  N}{k_c}} e^{-\phi_{II}(n,N)}dn 
    +\int_{\frac{\langle k \rangle  N}{ k_c}}^{N}
    e^{-\phi_{I}(n,N)}dn\right\}, 
\end{equation}
where subindices I and II refer to which type of potential is
dominating the integral. In the limit $N \gg 1$, we can evaluate the
contribution of each integral, leading to Eq.~(\ref{eq:13}).

\end{appendix}


\end{document}